\documentclass[a4paper,11pt]{article}
\pdfoutput=1 

\usepackage{jheppub} 
\usepackage[utf8]{inputenc}
\usepackage{graphicx}
\usepackage{appendix}

\usepackage{natbib}
\usepackage{slashed}
\usepackage{makecell}
\usepackage{bm}
\usepackage{lineno}
\usepackage{orcidlink}

\title{Dark radiation from Kerr primordial black holes: the role of superradiance}

\author[a]{Nayun Jia,}
\author[a,\ast]{Chen Zhang}\note[$\ast$]{Corresponding author.} 
\author[a,b,c,\ast]{and Xin Zhang}

\affiliation[a]{Liaoning Key Laboratory of Cosmology and Astrophysics, College of Sciences,
  Northeastern University, NO.~3-11, Wenhua Road, Heping District, Shenyang 110819, China}
\affiliation[b]{MOE Key Laboratory of Data Analytics and Optimization for Smart Industry,
  Northeastern University, NO.~3-11, Wenhua Road, Heping District, Shenyang 110819, China}
\affiliation[c]{National Frontiers Science Center for Industrial Intelligence and Systems Optimization,
  Northeastern University, NO.~3-11, Wenhua Road, Heping District, Shenyang 110819, China}

\emailAdd{nayun.jia@foxmail.com}
\emailAdd{zhangchen2@neu.edu.cn}
\emailAdd{zhangxin@neu.edu.cn}

\abstract{
    Light primordial black holes (PBHs) that fully evaporate before Big Bang Nucleosynthesis (BBN) produce dark radiation (DR) via Hawking radiation of gravitons, contributing to the effective number of relativistic species $\Delta N_{\rm eff}$.
    If the particle spectrum contains a beyond-the-Standard-Model (BSM) boson with Compton wavelength comparable to the black hole (BH) gravitational radius, superradiant instability extracts angular momentum from the BH into a bosonic cloud, whose gravitational wave (GW) emission contributes an additional source of DR.
    By simultaneously evolving the BH mass and spin, superradiant mode occupation numbers, comoving entropy and cosmological energy densities in an expanding early-universe background, we find that superradiance generically suppresses $\Delta N_{\rm eff}$: by extracting angular momentum before Hawking radiation can convert it into gravitons, superradiance starves the dominant dark-radiation channel.
    The GWs emitted by the superradiant cloud can partially compensate this loss, but only when the superradiant and BH evaporation timescales are comparable; otherwise the cloud GWs are emitted too early and diluted by cosmological expansion.
    For scalar bosons with gravitational coupling $\alpha_{\rm ini} = M_{\rm ini}\mu = 0.1$--$0.3$, with $M_{\rm ini}$ the initial PBH mass and $\mu$ the boson mass, superradiance pushes $\Delta N_{\rm eff}$ below the projected CMB-HD $2\sigma$ sensitivity ($\approx 0.027$) across the entire $(M_{\rm ini},\,a_{*,\rm ini})$ plane, closing the marginal detectability window that exists in the Hawking-only scenario.
    Vector bosons produce an even stronger suppression, as their faster superradiant growth leads to earlier GW emission and correspondingly greater redshift dilution.
    These results imply that existing $\Delta N_{\rm eff}$ bounds on PBH mass and spin derived without superradiance must be revisited if BSM bosons are present in the particle spectrum.
}

\begin{document}
\maketitle
\flushbottom

\section{Introduction}\label{sec:intro}

The early universe before Big Bang Nucleosynthesis (BBN) is one of the least constrained epochs in cosmology.
Unlike BBN or the Cosmic Microwave Background (CMB), which leave directly observable fossil records, the pre-BBN era offers no comparable relics: thermalization in the hot plasma efficiently erases most information about the prior history, leaving this epoch largely inaccessible to direct observation.

A rare exception arises from dark radiation (DR)---relativistic energy density beyond Standard Model (SM) photons and neutrinos~\cite{Kolb:1990vq,Ackerman:2008kmp,Archidiacono:2011gq}. Once produced, DR decouples from the SM plasma and does not participate in subsequent thermalization; its energy density is therefore frozen in and carried forward as a fossil of the epoch in which it was generated.
The imprint of DR is captured by the effective number of relativistic species, $N_{\mathrm{eff}}$, which is encoded in the CMB anisotropies~\cite{Planck:2018vyg} and in light-element abundances~\cite{Hamann:2010bk}. Any deviation $\Delta N_{\mathrm{eff}}$ from the SM prediction $N_{\mathrm{eff}}^{\rm SM} \simeq 3.044$~\cite{Mangano:2005cc,deSalas:2016ztq,Bennett:2019ewm,Bennett:2020zkv} thus provides one of the few observational windows into pre-BBN physics.
Next-generation CMB experiments are poised to sharpen this window considerably: for example, CMB-HD~\cite{Sehgal:2019ewc,MacInnis:2023vif} targets a $2\sigma$ sensitivity of $\Delta N_{\mathrm{eff}} \approx 0.027$, bringing previously inaccessible shifts within observational reach.

Primordial black holes (PBHs; for comprehensive reviews, see refs.~\cite{Carr:2020gox,Auffinger:2022khh,Escriva:2022duf}) offer a mechanism for generating DR in the early universe.
Formed from large primordial density fluctuations, PBHs could have been present well before BBN and may even have temporarily dominated the cosmic energy budget~\cite{Zeldovich:1967lct,Hawking:1971ei,Carr:1974nx,Carr:1975qj,Sasaki:2018dmp,Hooper:2019gtx}.
Through Hawking radiation~\cite{Hawking:1975vcx}, PBHs can efficiently emit all particle species with masses below the Hawking temperature, regardless of how weakly they couple to the SM, and consequently inject both SM particles and any additional relativistic degrees of freedom into the cosmic plasma.
This universal particle production makes PBHs a compelling source of DR whose contribution to $\Delta N_{\mathrm{eff}}$ warrants precise quantification.
In this work, we focus on PBHs that are light enough to fully evaporate before BBN, so as not to conflict with BBN constraints~\cite{Carr:2009jm,Wu:2025ovd}.\footnote{For PBHs that survive into later epochs, other probes become relevant, such as the 21~cm signal~\cite{Zhao:2025ekx,Zhao:2025ddy,Cang:2021owu} and gamma-ray bursts from the final evaporation of PBHs with initial masses around $10^{15}~\mathrm{g}$~\cite{LHAASO:2025kyn,Yang:2024vij,Yang:2025uvf}.}

The spin of a PBH directly affects its emission spectrum and efficiency.
For non-rotating (Schwarzschild) black holes (BHs), greybody factors suppress the Hawking radiation of higher-spin fields, making spin-1 and spin-2 states subdominant in the emission spectrum~\cite{Page:1976df}.
This is no longer the case for rotating (Kerr) BHs: a Kerr BH radiates vector bosons and gravitons far more efficiently~\cite{Page:1976ki}, which can shift $\Delta N_{\mathrm{eff}}$ by a phenomenologically relevant amount~\cite{Arbey:2021ysg,Masina:2021zpu,Hooper:2020evu}.
However, the angular momentum of a Kerr PBH is depleted much more rapidly than its mass, so the spin-enhanced graviton emission is concentrated in the early stages of Hawking radiation.
Particles radiated at these early times experience more cosmological redshift, partially offsetting the enhancement.
A consistent treatment that simultaneously tracks PBH mass loss, spin evolution, and the expansion of the universe shows that neglecting this redshift suppression can overestimate $\Delta N_{\mathrm{eff}}$ by up to an order of magnitude for near-extremal PBHs~\cite{Cheek:2022dbx}, underscoring the necessity of precise modeling for reliable predictions.

Beyond Hawking radiation, rotating BHs can undergo superradiance\footnote{Throughout this work, ``superradiance'' refers specifically to the \emph{superradiant instability}, a spontaneous and sustaining process in which bosonic quasi-bound states grow exponentially around a Kerr BH, as opposed to the one-time amplification of scattered waves known as \emph{superradiant scattering}.} (for a comprehensive review, see ref.~\cite{Brito:2015oca}) if the particle species contain a boson whose Compton wavelength is comparable to the BH gravitational radius.
Superradiance continuously extracts rotational energy and angular momentum from the BH, building up a macroscopic bosonic cloud.
This process can significantly accelerate the spin-down of PBHs, storing a sizable fraction of the rotational energy in the cloud.
The cloud subsequently dissipates through gravitational wave (GW) emission, converting this stored energy into gravitational radiation.
In a cosmological setting, such gravitational radiation represents a contribution to DR and is qualitatively distinct from the gravitons directly emitted via Hawking radiation.

Existing analyses of DR from Kerr PBHs have incorporated spin-dependent Hawking radiation and cosmological redshift~\cite{Cheek:2022dbx}, while
the role of superradiance has been explored in the context of dark matter production from PBHs~\cite{Jia:2025vqn,Bernal:2021bbv} and ALP dark radiation from moduli amplified by superradiance~\cite{Manno:2025dhw}.
It is natural to ask how superradiance modifies the DR yield.
In this work, we address this question by tracking the joint evolution of PBH mass and spin under both Hawking radiation and superradiance, within an expanding cosmological background that consistently accounts for the redshift of emitted particles.
In particular, we examine how the GWs radiated by the superradiant cloud contribute to DR, and we assess the resulting implications for $\Delta N_{\mathrm{eff}}$.

The combined effect on $\Delta N_{\mathrm{eff}}$ is not a naive sum of the two contributions. Both Hawking radiation and superradiance act on the same dynamical variables---BH mass and spin: superradiance accelerates BH spin-down, which consequently reduces the efficiency of spin-enhanced Hawking radiation, while the timescale of superradiance relative to BH lifetime determines the extent to which the superradiant cloud GWs are cosmologically redshifted.
This interplay makes the net effect on $\Delta N_{\mathrm{eff}}$ hard to assess without solving the full system numerically.
In this work, we find that superradiance generically suppresses $\Delta N_{\mathrm{eff}}$ relative to the Hawking radiation-only baseline.
By extracting angular momentum before Hawking radiation can convert it into gravitons, superradiance reduces the Hawking DR channel, and the superradiant cloud GWs can only partially compensate the loss.
At the phenomenological level, superradiance closes the $\Delta N_{\mathrm{eff}}$ detectability window that CMB-HD would otherwise have for near-extremal Kerr PBHs.

The rest of this paper is organized as follows.
In sec.~\ref{sec:kerr_pbh}, we briefly review the formation scenarios of PBHs and the standard formalism of Hawking radiation for Kerr BHs, establishing the baseline for particle production in the absence of superradiance.
In sec.~\ref{sec:superradiance}, we detail the physics of superradiant instability, incorporating the effects of multiple modes and the GW emission from the superradiant cloud.
Sec.~\ref{sec:evolution_neff} presents the definition of $\Delta N_{\mathrm{eff}}$ and its observational context, the comoving entropy evolution and SM heating, and the complete coupled system of evolution equations.
We present our results in sec.~\ref{sec:results}, and summarize our findings and discuss their implications in sec.~\ref{sec:conclusions}.

In this paper, we use natural units where $G=\hbar=c=k_{\mathrm{B}}=1$.

\section{Kerr primordial black holes in the early universe}\label{sec:kerr_pbh}

\subsection{Formation and initial conditions}

In the standard picture, PBHs form from the gravitational collapse of overdense regions when the associated length scale re-enters the Hubble horizon;
alternative formation channels include bubble collisions and collapse of delayed false-vacuum patches during first-order phase transitions~\cite{Hawking:1982ga,Jung:2021mku,Liu:2021svg,Cai:2024nln,Ning:2026nfs,Kanemura:2024pae}, as well as domain-wall collapse during second-order phase transitions~\cite{Rubin:2000dq,Deng:2017uwc,Liu:2019lul,Lu:2024ngi,Lu:2024szr}.
A cosmological PBH population is specified by its mass and spin distributions, and overall abundance. In this work, we consider PBH formation during the radiation-dominated era after reheating and adopt a monochromatic mass function, so that all PBHs form with a single initial mass $M_{\rm ini}$.\footnote{For extended mass functions, see, e.g., ref.~\cite{Carr:2016drx}.} This mass scales with the horizon mass at the formation cosmic temperature $T_{\rm ini}$,
\begin{equation}
    M_{\rm ini} = \frac{4\pi}{3}\,\gamma\,\frac{\rho_R(T_{\rm ini})}{H^3(T_{\rm ini})}\,,
    \label{eq:Mini}
\end{equation}
where $\gamma \approx 0.2$ is a gravitational collapse efficiency factor~\cite{Carr:1975qj,Choptuik:1992jv,Musco:2012au} and $H(T_{\rm ini})$ is the Hubble parameter at formation.
Through the standard Friedmann equation in radiation domination, this implies $T_{\rm ini}\propto M_{\rm ini}^{-1/2}$.
CMB constraints on the inflationary Hubble parameter set a lower bound $M_{\rm ini}\gtrsim 0.1\,\mathrm{g}$~\cite{Green:1997sz,Planck:2018jri}.
We additionally require complete PBH evaporation before the onset of BBN, which restricts our parameter space to $M_{\rm ini}\lesssim 10^{9}\,\mathrm{g}$~\cite{Carr:2009jm,Wu:2025ovd}.

PBHs can acquire angular momentum, which we denote by $J$, through various mechanisms. During formation, angular momentum may arise from processes such as domain-wall collapse~\cite{Rubin:2000dq}, gravitational collapse during a matter-dominated era~\cite{Harada:2017fjm}, or assembly of compact objects such as Q-balls~\cite{Cotner:2016cvr} or oscillons~\cite{Cotner:2017tir,Cotner:2018vug,Flores:2021tmc}. After formation, PBHs can also spin up due to the emission of light scalar particles~\cite{Calza:2021czr,Calza:2023rjt}.
A convenient measure is the dimensionless spin parameter $a_\ast \equiv J/M^2 \in [0,1)$.
For PBHs formed via the standard mechanism during radiation domination, the initial spin is typically estimated to be small, $a_{\ast}\sim\mathcal{O}(10^{-3})$~\cite{Chiba:2017rvs,DeLuca:2019buf,Mirbabayi:2019uph,He:2019cdb,Harada:2020pzb}, though alternative formation channels may yield substantially higher values~\cite{Banerjee:2023qya}.
In this work, we remain agnostic about the specific PBH spin generation mechanism and treat the initial spin $a_{\ast,\mathrm{ini}}$ as a free parameter.

The overall PBH abundance at formation is expressed in the initial energy density fraction
\begin{equation}
    \beta \equiv \left.\frac{\rho_{\rm PBH}}{\rho_{\rm R}}\right|_{t=t_{\rm ini}}\,,
    \label{eq:beta_def}
\end{equation}
where $\rho_{\rm PBH}$ is the PBH energy density, $\rho_{\rm R}$ the Standard-Model radiation energy density, and $t_{\rm ini}$ the formation time.
As the universe expands, $\rho_{\rm PBH}\propto a^{-3}$ while $\rho_R\propto a^{-4}$, where $a$ is the cosmological scale factor\footnote{Not to be confused with the dimensionless spin parameter $a_\ast$ introduced above.}, so the ratio $\rho_{\rm PBH}/\rho_R$ grows linearly with $a$.
If the PBHs are sufficiently long-lived, they will temporarily dominate the energy budget, giving rise to an early matter-dominated era.
This occurs when $\beta$ exceeds a critical value
\begin{equation}
    \beta_c \approx 6.4\times 10^{-10}\left(\frac{M_{\rm ini}}{10^4\,\mathrm{g}}\right)^{-1}\,.
    \label{eq:beta_c}
\end{equation}
Whether PBH domination is achieved has important consequences for the resulting DR yield, as we will discuss in sec.~\ref{sec:evolution_neff}.

\subsection{Kerr black hole and Hawking radiation}

For a Kerr BH with mass $M$ and angular momentum $J \equiv M^2 a_\ast$, the metric in Boyer-Lindquist coordinates $(t, r, \theta, \varphi)$ reads~\cite{Kerr:1963ud,Boyer:1966qh,Teukolsky:1973ha}
\begin{equation}
    \begin{aligned}
        \mathrm{d} s^2= & -\mathrm{d} t^2+\frac{2 M r}{\Sigma}\left(\mathrm{~d} t-M a_\ast \sin ^2 \theta \mathrm{~d} \varphi\right)^2                        \\
                        & +\frac{\Sigma}{\Delta} \mathrm{d} r^2+\Sigma \mathrm{d} \theta^2+\left(r^2+M^2 a_\ast^2\right) \sin^2 \theta \mathrm{~d} \varphi^2,
    \end{aligned}
\end{equation}
where $\Sigma \equiv r^2+M^2 a_\ast^2 \cos ^2 \theta$ and $\Delta \equiv r^2-2 M r+M^2 a_\ast^2$.
The Hawking radiation spectrum of a Kerr BH depends on several horizon quantities.
The outer horizon radius is
\begin{equation}
    r_+ = M\left(1+ \sqrt{1-a_\ast^2}\right)\,.
\end{equation}
In terms of $r_+$, the angular velocity and Hawking temperature of the outer horizon are
\begin{align}
    \Omega_H & = \frac{a_\ast}{2\,r_+}\,,               \\
    T_H      & = \frac{\sqrt{1-a_\ast^2}}{4\pi\,r_+}\,,
\end{align}
with the Schwarzschild limit $T_H=1/(8\pi M)$ recovered as $a_\ast\to 0$.

The particle emission rate from a Kerr BH is mode-dependent and is shaped by greybody factors (transmission probabilities through the gravitational potential barrier)~\cite{Hawking:1975vcx,Page:1976df,Page:1976ki}.
For a particle species $i$ with spin $s_i$ and internal degrees of freedom $g_i$, the differential emission rate is
\begin{equation}
    \frac{\mathrm{d}^2 N^\mathrm{HR}_i}{\mathrm{~d} \omega \mathrm{~d} t}=\frac{g_i}{2 \pi} \sum_{l=s_i} \sum_{m=-l}^l \frac{\mathrm{~d}^2 N_{i, l m}}{\mathrm{~d} \omega \mathrm{~d} t}
\end{equation}
with
\begin{equation}
    \frac{\mathrm{d}^2 N_{i, l m}}{\mathrm{~d} \omega \mathrm{~d} t}=\frac{\Gamma_{s_i}^{l m}\left(M_{\mathrm{BH}}, \omega, a_\ast\right)}{\exp \left[\left(\omega-m \Omega_{\mathrm{H}}\right) / T_{\mathrm{BH}}\right]-(-1)^{2 s_i}},
\end{equation}
where $\omega$ is the particle energy, $(l,m)$ label the spheroidal harmonics, and $\Gamma_{s_i}^{l m}$ is the greybody factor.
The factor $(-1)^{2s}$ equals $+1$ for bosons and $-1$ for fermions, reproducing the Bose--Einstein and Fermi--Dirac statistics, respectively.
The quantity $\omega - m\Omega_H$ is the mode frequency with respect to the horizon-generating Killing vector $\chi = \partial_t + \Omega_H\,\partial_\phi$, and $\omega<m\Omega_H$ is the superradiance condition that will be discussed in sec.~\ref{sec:superradiance}.

Integrating the differential rate over all energies gives the total particle number emission rate for species $i$,
\begin{equation}
    \Gamma^{\rm HR}_i (M,a_\ast) \equiv \int_0^\infty {\rm d}\omega\;\frac{\mathrm{d}^2 N^\mathrm{HR}_i}{\mathrm{d}\omega\,\mathrm{d}t}\,,
    \label{eq:Gamma_HR_i}
\end{equation}
which counts the number of particles of species $i$ emitted per unit time.

The total BH mass and angular-momentum loss rates follow from weighted integrals of the spectrum, summed over all particle species,
\begin{align}
    \frac{{\rm d}M}{{\rm d}t}\bigg|_{\rm HR} & =-\sum_{i}\int_0^\infty {\rm d}\omega\;\omega\,
    \frac{\mathrm{d}^2 N^\mathrm{HR}_i}{\mathrm{~d} \omega \mathrm{~d} t}\,,\label{eq:dMdt_HR} \\
    \frac{{\rm d}J}{{\rm d}t}\bigg|_{\rm HR} & =-\sum_{i}\int_0^\infty {\rm d}\omega\;m\,
    \frac{\mathrm{d}^2 N^\mathrm{HR}_i}{\mathrm{~d} \omega \mathrm{~d} t}\,.\label{eq:dJdt_HR}
\end{align}
In practice, it is conventional to define the Page functions~\cite{Page:1976df,Page:1976ki}
\begin{align}
    f(M,a_\ast) & \equiv M^2\,\sum_{i}\int_0^\infty {\rm d}\omega\;\omega\,
    \frac{\mathrm{d}^2 N^\mathrm{HR}_i}{\mathrm{~d} \omega \mathrm{~d} t}\,,\label{eq:page_f} \\
    g(M,a_\ast) & \equiv \frac{M}{a_\ast}\,\sum_{i}\int_0^\infty {\rm d}\omega\;m\,
    \frac{\mathrm{d}^2 N^\mathrm{HR}_i}{\mathrm{~d} \omega \mathrm{~d} t}\,.\label{eq:page_g}
\end{align}
Due to the dependence on BH spin and emitted particle content, accurately determining the Page functions is technically nontrivial. In this paper, we use numerically computed Page functions from open-source tools \texttt{BlackHawk}~\cite{Arbey:2019mbc, Arbey:2021mbl} and \texttt{FRISBHEE}~\cite{Cheek:2021odj, Cheek:2021cfe, Cheek:2022dbx, Cheek:2022mmy} to capture the full dynamics of Hawking evaporation.

In terms of the Page functions, the loss rates for $M$ and $a_\ast$ can take the compact form
\begin{align}
    \frac{{\rm d}M}{{\rm d}t}\bigg|_{\rm HR}\left(M,a_\ast\right)      & =-\frac{f(M,a_\ast)}{M^2}\,,\label{eq:dMdt_page}                   \\
    \frac{{\rm d}a_\ast}{{\rm d}t}\bigg|_{\rm HR}\left(M,a_\ast\right) & =a_\ast\frac{2f(M,a_\ast)-g(M,a_\ast)}{M^3}\,.\label{eq:dadt_page}
\end{align}
These equations will be augmented by the superradiant contributions in sec.~\ref{sec:superradiance} and embedded in the full cosmological evolution in sec.~\ref{sec:evolution_neff}.

\subsection{Dark radiation from Hawking radiation}

Hawking radiation universally produces all particle species with masses below the Hawking temperature $T_H$, regardless of their coupling to the SM.
In particular, gravitons are always emitted~\cite{Anantua:2008am,Dong:2015yjs,Hooper:2020evu,Barman:2024ufm}, and since they do not thermalize with the SM plasma, they propagate as free-streaming radiation from the moment of production, and thus contribute to DR energy density.

Kerr rotation enhances emission into higher-spin states, leading to a relatively rapid spin-down phase followed by a Schwarzschild stage.
This can be made explicit by defining the characteristic timescales
\begin{equation}
    \begin{aligned}
        \tau_M   & \equiv M/\left|\dot M\right|=M^3/f \ ,               \\
        \tau_{a} & \equiv a_\ast/\left|\dot a_\ast\right|=M^3/(g-2f)\ .
    \end{aligned}
\end{equation}
We also define $\tau_{\rm BH}$ as the total BH lifetime evaluated at the initial mass $M_{\rm ini}$.
For the SM particle content plus gravitons, numerical calculations show that the BH spin is always significantly reduced well before full evaporation~\cite{Page:1976df,Page:1976ki,Taylor:1998dk}, i.e.,
\begin{equation}
    \frac{\tau_{a}}{\tau_M}=\frac{f}{g-2f} \ll 1\,.\label{eq:timescale_ratio}
\end{equation}
This conclusion depends on the spin structure of the radiated species: if the particle spectrum is dominated by scalar ($s=0$) fields, the preferential energy loss can slow, stall, or even reverse the spin-down~\cite{Chambers:1997ai,Taylor:1998dk,Calza:2021czr,Calza:2023rjt}.
For the SM spectrum considered in this work, however, the particle spectrum ensures that the spin-down is always faster.

The Kerr-enhanced graviton emission---and hence the excess DR production relative to a Schwarzschild PBH---is therefore confined to this brief early window.
Because these gravitons are emitted well before evaporation completes, they experience correspondingly more cosmological redshift.
Quantifying this suppression requires tracking the emission jointly with the expansion history, which we carry out in sec.~\ref{sec:evolution_neff}.

A further effect that modifies the observable DR yield is entropy injection into the SM plasma.
As PBHs evaporate, their energy density is converted into relativistic particles, dominantly SM species. This process increases the comoving SM entropy
\begin{equation}
    \mathcal{S}\equiv s\,a^3\,,\label{eq:com_entropy}
\end{equation}
and the SM entropy density is
\begin{equation}
    s(T)=\frac{2\pi^2}{45}\,g_{*S}(T)\,T^3\,.\label{eq:entropy_density}
\end{equation}
Here $g_{*S}(T)$ is the effective number of entropic degrees of freedom, defined by
\begin{equation}
    g_{*S}(T)\equiv \sum_{\text{bosons}}g_i\!\left(\frac{T_i}{T}\right)^{\!3}+\frac{7}{8}\sum_{\text{fermions}}g_i\!\left(\frac{T_i}{T}\right)^{\!3}\,,\label{eq:gstar_s}
\end{equation}
where $T_i$ is the temperature of species $i$ (for species in thermal equilibrium with the SM plasma, $T_i=T$).
The counterpart of $g_{*S}$ for energy density, i.e., the effective number of relativistic degrees of freedom $g_*(T)$, is defined by
\begin{equation}
    g_*(T)\equiv \sum_{\text{bosons}}g_i\!\left(\frac{T_i}{T}\right)^{\!4}+\frac{7}{8}\sum_{\text{fermions}}g_i\!\left(\frac{T_i}{T}\right)^{\!4}\,.\label{eq:gstar}
\end{equation}

Because the DR sector is already decoupled from the SM at the epoch of PBH evaporation, it does not participate in the entropy injection. Its comoving energy density is therefore conserved,
\begin{equation}
    \rho_{\rm DR}\,a^4=\text{const}\,.\label{eq:rho_DR_conserved}
\end{equation}
By contrast, the SM radiation temperature at a fixed scale factor is set by the comoving SM entropy. From eqs.~\eqref{eq:com_entropy}--\eqref{eq:entropy_density}, $s=\mathcal{S}/a^3$, so
\begin{equation}
    T=\left(\frac{45\,\mathcal{S}}{2\pi^2\,g_{*S}(T)\,a^3}\right)^{\!1/3}\,.
\end{equation}
Since the SM radiation energy density is $\rho_{\rm SM}=\frac{\pi^2}{30}\,g_*(T)\,T^4$, we obtain
\begin{equation}
    \rho_{\rm SM}
    = \frac{3}{4}\!\left(\frac{45}{2\pi^2}\right)^{\!1/3}
    g_*(T)\,g_{*S}(T)^{-4/3}\,\mathcal{S}^{4/3}\,a^{-4}\,.
    \label{eq:rho_SM_scaling}
\end{equation}
As discussed above, the BH angular momentum is depleted on a timescale $\tau_a\ll\tau_M$, after which the graviton emission rate falls to a minimal value.
Yet the BH continues to radiate mainly SM species for the remainder of its lifetime, with a growing radiation power due to the increasing Hawking temperature.
This continued emission significantly increases $\mathcal{S}$, whereas the gravitons, already decoupled and now emitted only negligibly, simply redshift according to eq.~\eqref{eq:rho_DR_conserved}.
This dilution acts as a late-time suppression on DR, partially counteracting the early Kerr enhancement.

The interplay between spin-enhanced emission, cosmological redshift, and entropy injection defines the baseline DR yield from Hawking radiation alone. In the following sections, we will show that superradiance introduces a distinct production channel (sec.~\ref{sec:superradiance}), and track both contributions simultaneously within an expanding early universe (sec.~\ref{sec:evolution_neff}).

\section{Superradiant instability and gravitational wave emission}\label{sec:superradiance}

\subsection{Quasi-bound states and superradiance condition}

A Kerr BH can spontaneously develop a superradiant instability~\cite{Press:1972zz,Damour:1976kh,Zouros:1979iw,Detweiler:1980uk} whenever there exists a beyond-the-Standard-Model (BSM) bosonic species with mass $\mu$ whose Compton wavelength $\sim1/\mu$ is comparable to the BH gravitational radius $\sim M$.
The instability extracts energy and angular momentum from the BH and builds up a macroscopic bosonic cloud in its vicinity.
The relevant dimensionless gravitational coupling is
\begin{equation}
    \alpha \equiv M\mu\,,
    \label{eq:alpha_def}
\end{equation}
and efficient superradiance requires $\alpha\lesssim\mathcal{O}(1)$.

The Kerr gravitational potential supports hydrogen-like quasi-bound states~\cite{Arvanitaki:2009fg,Arvanitaki:2010sy,Baumann:2019eav}.
For a scalar ($s=0$) boson, these states are labeled by $|nlm\rangle$, i.e., the principal quantum number $n$, the orbital angular momentum quantum number $l$, and the azimuthal quantum number $m$, subject to $0 \leq l \leq n-1$ and $-l\leq m\leq l$.
For a massive vector ($s=1$) boson, the states carry an additional total angular momentum quantum number $j$ and are denoted $|nljm\rangle$.\footnote{In general, $j$ can take values $j=l \pm 1$ (the so-called ``electric modes") or $j=l$ (the ``magnetic modes"). Throughout this paper we restrict our discussion to $j=l+1$. A physical interpretation of the $j$--$l$ relation can be found in ref.~\cite{Baumann:2019eav}.}
Each quasi-bound state possesses a complex eigenfrequency $\omega = \omega_R + i\,\omega_I$. In the hydrogenic regime $\alpha\ll 1$, the real part takes the form
\begin{equation}
    \omega_R \simeq \mu\!\left(1-\frac{\alpha^2}{2n^2}\right),
    \label{eq:omega_R}
\end{equation}
while the imaginary part $\omega_I$ governs the instability of the mode~\cite{Dolan:2007mj}.\footnote{For phenomenological purposes, note that $\omega_R < \mu$ but the two differ only by $\mathcal{O}(\alpha^2)$. Throughout this work, we therefore approximate $\omega_R \approx \mu$ when evaluating the energy per superradiant particle in the cloud, reducing computational complexity without fundamentally changing the results. For the precise value of $\omega_R$, see ref.~\cite{Baumann:2019eav}.}

A mode grows exponentially when the superradiance condition
\begin{equation}
    \omega_R < m\,\Omega_H
    \label{eq:sr_condition}
\end{equation}
is satisfied, yielding $\omega_I > 0$.
This is the same condition identified in sec.~\ref{sec:kerr_pbh}.
For quasi-bound states, superradiance is self-reinforcing and drives exponential growth of a macroscopic bosonic cloud.

As the cloud grows and extracts BH angular momentum, $\Omega_H$ decreases until the condition~\eqref{eq:sr_condition} ceases to hold. For each mode, this defines a critical spin below which growth stops,
\begin{equation}
    a_{\ast c} \simeq \frac{4m\,\alpha}{m^2+4 \alpha^2}\,.
    \label{eq:a_crit}
\end{equation}
As discussed in sec.~\ref{sec:kerr_pbh}, Hawking radiation of gravitons carries away angular momentum on its own characteristic timescale. The two processes therefore compete, and the resulting evolution is governed by their relative timescales.

\subsection{Mode growth and BH--cloud energy transfer}

The superradiant instability rate $\Gamma^{\rm SR} \equiv 2\,\omega_{I}$ determines the timescale on which the cloud forms. For a scalar mode $|nl m\rangle$ in the regime $\alpha\ll 1$, the superradiance rate calculated from matched asymptotic expansion reads~\cite{Brito:2015oca,Baumann:2019eav}
\begin{equation}
    \begin{aligned}
        \Gamma^{\rm SR}_{nl m}= & \frac{4r_+}{M}\,C_{nl}(m\Omega_H-\omega_R)\,\alpha^{4l+5}                         \\
                                & \times \prod_{k=1}^{l}\!\left[k^2(1-a_\ast^2)+(a_\ast m-2r_+\omega_R)^2\right]\,,
    \end{aligned}\label{eq:Gamma_sr_scalar}
\end{equation}
where
\begin{equation}
    C_{n l} \equiv \frac{2^{4 l+1}(n+l)!}{n^{2 l+4}(n-l-1)!}\left[\frac{l!}{(2 l)!(2 l+1)!}\right]^2
\end{equation}
is a mode-dependent numerical coefficient. For a vector mode $|nl jm\rangle$, the corresponding rate reads
\begin{equation}
    \begin{aligned}
        \Gamma^{\rm SR}_{nl jm}= & \frac{4r_+}{M}\,C_{nl j}(m\Omega_H-\omega_R)\,\alpha^{2l+2j+5}                  \\
                                 & \times \prod_{k=1}^{j}\!\left[k^2(1-a_\ast^2)+(a_\ast m-2r_+\omega_R)^2\right],
    \end{aligned}\label{eq:Gamma_sr_vector}
\end{equation}
with
\begin{equation}
    \begin{aligned}
        C_{n l j} \equiv & \frac{2^{2 l+2 j+1}(n+l)!}{n^{2 l+4}(n-l-1)!}\left[\frac{(l)!}{(l+j)!(l+j+1)!}\right]^2 \\
                         & \times\left[1+\frac{2(1+l-j)(1-l+j)}{l+j}\right]^2 .
    \end{aligned}
\end{equation}
The steep $\alpha$-dependence favors modes with the lowest possible $l$. Among scalar modes, $|211\rangle$ ($l=1$, scaling as $\alpha^9$) grows fastest~\cite{Dolan:2007mj}; among vector modes, $|1011\rangle$ ($l=0,\,j=1$, scaling as $\alpha^7$) is the dominant one~\cite{Baryakhtar:2017ngi,East:2017ovw,Jia:2023see}.

In the linear regime, the occupation number $N^\mathrm{SR}_i$ of each superradiant mode evolves as
\begin{equation}
    \frac{{\rm d}N^\mathrm{SR}_i}{{\rm d}t}=\Gamma^{\rm SR}_i\,N^\mathrm{SR}_i\,,
    \label{eq:dNdt_sr}
\end{equation}
seeded by quantum fluctuations at a small initial occupation number, whose value does not essentially change the mode evolution~\cite{Arvanitaki:2010sy,Bernal:2022oha}. The energy and angular momentum stored in the cloud of mode $i$ are
\begin{equation}
    M^\mathrm{SR}_{i}=\mu \,N^\mathrm{SR}_i\,,\qquad J^\mathrm{SR}_{i}=m_i\,N^\mathrm{SR}_i\,.
    \label{eq:cloud_EL}
\end{equation}

As the cloud grows, the BH loses mass and angular momentum at the rates
\begin{align}
    \left.\frac{{\rm d}M}{{\rm d}t}\right|_{\rm SR} & =-\sum_i\Gamma^{\rm SR}_i M^\mathrm{SR}_{i}\,,\label{eq:dMdt_sr}    \\
    \left.\frac{{\rm d}J}{{\rm d}t}\right|_{\rm SR} & =-\sum_i \,\Gamma^{\rm SR}_i J^\mathrm{SR}_{i}\,.\label{eq:dJdt_sr}
\end{align}
In terms of the dimensionless spin parameter, the superradiant contribution to the spin evolution is
\begin{equation}
    \left.\frac{{\rm d}a_\ast}{{\rm d}t}\right|_{\rm SR}=\sum_i\!\left(\frac{2a_\ast\,\mu}{M}-\frac{m_i}{M^2}\right)\Gamma^{\rm SR}_i\,N^\mathrm{SR}_i\,.
    \label{eq:dadt_sr}
\end{equation}
These are to be combined with the Hawking-radiation loss rates in eqs.~\eqref{eq:dMdt_page}--\eqref{eq:dadt_page}.

When multiple modes are simultaneously superradiant, the mode with the shortest $\tau_{\rm SR}$ dominates initially. As this primary mode extracts BH angular momentum toward its $a_{\ast c}$, its growth stalls, and a secondary mode with a lower $a_{\ast c}$ can subsequently take over. Following ref.~\cite{Jia:2025vqn}, we include two modes per boson species: the $|211\rangle$ and $|322\rangle$ modes for scalar bosons, and the $|1011\rangle$ and $|2122\rangle$ modes for vector bosons. This two-mode treatment captures the essential sequential dynamics across the PBH parameter space of interest.

\subsection{GW emission and cloud dissipation}\label{sec:GW_cloud}

The superradiant cloud dissipates through GW emission, which contributes to DR that is qualitatively distinct from the Hawking-radiated gravitons discussed in sec.~\ref{sec:kerr_pbh}.
The superradiant GW channel corresponds to the annihilation of boson pairs within the same mode into gravitons.\footnote{Gravitons can also be produced via cross-mode effects. However, the powers of these channels are suppressed. See appendix B of ref.~\cite{Jia:2025vqn} and references therein.}
This process produces monochromatic GWs with frequency
\begin{equation}
    \omega_{\rm GW} = 2\,\omega_R \approx 2\mu\,.
    \label{eq:omega_GW}
\end{equation}
For a bosonic mode $i$, the GW power from annihilation takes the form~\cite{Brito:2015oca, Guo:2022mpr, Guo:2024dqd, Luo:2026eef}
\begin{equation}
    P^{\rm GW}_i=\mathcal{C}_i\,\left(\frac{{M^\mathrm{SR}_i}}{M}\right)^2\,\alpha^{4l+10}\,,
    \label{eq:PGW_scalar}
\end{equation}
where $\mathcal{C}_i$ is a mode-dependent numerical coefficient.
The GW emission depletes the cloud occupation number at the rate
\begin{equation}
    \left.\frac{{\rm d}N^\mathrm{SR}_i}{{\rm d}t}\right|_{\rm GW}=-\frac{P^{\rm GW}_i}{\mu}\,,
    \label{eq:dNdt_GW}
\end{equation}

Combining the superradiant growth~\eqref{eq:dNdt_sr} with GW depletion~\eqref{eq:dNdt_GW}, the full occupation-number evolution for each mode is
\begin{equation}
    \frac{{\rm d}N^\mathrm{SR}_i}{{\rm d}t}=\Gamma^{\rm SR}_i\,N^\mathrm{SR}_i-\frac{P^{\rm GW}_i}{\mu}\,,
    \label{eq:dNdt_full}
\end{equation}
with the associated GW timescale $\tau^{{\rm GW}}_i$ defined by the decay pattern
\begin{equation}
    M^{\mathrm{SR}}_i(t)=\frac{M_{i,\max}^{\mathrm{SR}}}{1+t / \tau^{\mathrm{GW}}_i}.
    \label{eq:tau_GW}
\end{equation}
Therefore, for each mode in the superradiant cloud, it first grows exponentially on the superradiant timescale $\tau^{\rm SR}_i\equiv(\Gamma^{\rm SR}_i)^{-1}$, saturates at $M^{\rm SR}_{i, \max}$ once its superradiance condition is no longer satisfied, and is subsequently drained by GW emission on $\tau^{\rm GW}_i$.

Beyond GW emission, the cloud is subject to two additional drain channels.
The first is superradiant decay: once $a_\ast$ drops below $a_{\ast c}$, the condition $\omega_R < m\Omega_H$ is violated, $\Gamma^{\rm SR}_i$ changes sign, and the BH re-absorbs the cloud bosons on timescale $|\Gamma^{\rm SR}_i|^{-1}$~\cite{Brito:2015oca,Ficarra:2018rfu}.
In most cases, this is the dominant mechanism by which the superradiant cloud disappears.
The second is intrinsic decay of the BSM boson into SM species.
A BSM boson would behave as dark matter and risk overclosing the universe~\cite{Preskill:1982cy,Abbott:1982af,Arvanitaki:2009fg}, so we require $\Gamma^{\rm decay}_{\rm BSM}>0$.
In this work, $\Gamma^{\rm decay}_{\rm BSM}$ is chosen so that $\tau_{\rm decay}\equiv(\Gamma^{\rm decay}_{\rm BSM})^{-1}$ is long enough for the cloud to build up fully, yet short enough that the BSM boson does not survive to violate constraints on dark matter abundance.
Since superradiant decay dominates cloud depletion in most cases, this intrinsic decay channel mainly governs the lifetime of the Hawking-produced BSM boson population $N_{\rm BSM}^{\rm HR}$ rather than the cloud itself.
The full occupation-number equation incorporating all three channels is presented in sec.~\ref{sec:evolution_eqs}.

The ordering of the characteristic timescales $\tau_{\rm SR}\ll\tau_{\rm GW}\ll\tau_{\rm BH}$~\cite{Arvanitaki:2010sy,Brito:2015oca,East:2017mrj} frames the physical picture of this paper. The cloud builds up on $\tau_{\rm SR}$ before appreciable GW power has been radiated, and subsequently dissipates primarily through superradiant decay. GW emission continues on the longer timescale $\tau_{\rm GW}$ as a subdominant but DR-relevant channel. The entire superradiant episode completes well before the PBH has substantially evaporated; because superradiance extracts BH mass and angular momentum at early times, it reshapes the mass and spin available for subsequent Hawking production of gravitons.

The full cosmological evolution that tracks this combined DR yield is developed in sec.~\ref{sec:evolution_neff}.

\section{\texorpdfstring{$\Delta N_\mathrm{eff}$}{Delta Neff} and evolution equations}\label{sec:evolution_neff}

\subsection{\texorpdfstring{$\Delta N_{\mathrm{eff}}$}{Delta Neff}: definition and observational context}

In the radiation-dominated era, after $e^\pm$ annihilation, the total radiation energy density of the universe is conventionally parametrized as~\cite{Lesgourgues:2006nd,Dolgov:2002wy}
\begin{equation}
    \rho_{\rm rad} = \left[1 + \frac{7}{8}\left(\frac{4}{11}\right)^{4/3} N_{\rm eff}\right]\rho_\gamma\,,
    \label{eq:Neff_param}
\end{equation}
where $\rho_\gamma$ is the photon energy density and $N_{\rm eff}$ is the effective number of relativistic species.
The factor $(4/11)^{4/3}$ is from the neutrino-to-photon temperature ratio $(T_\nu/T_\gamma)=(4/11)^{1/3}$ established after $e^\pm$ annihilation.
In the SM, the three active neutrino species yield $N_{\rm eff}^{\rm SM} \simeq 3.044$~\cite{Mangano:2005cc,deSalas:2016ztq,Bennett:2019ewm,Bennett:2020zkv}, where the small departure from~3 reflects non-instantaneous neutrino decoupling and finite-temperature QED corrections.
Current constraints from Planck CMB anisotropies~\cite{Planck:2018vyg} and BBN light-element abundances~\cite{Hamann:2010bk} are consistent with this prediction.

Any deviation from the SM prediction,
\begin{equation}
    \Delta N_{\rm eff} \equiv N_{\rm eff} - N_{\rm eff}^{\rm SM}\,,
    \label{eq:Delta_Neff_def}
\end{equation}
signals the presence of additional relativistic energy density beyond SM photons and neutrinos. Decomposing the total radiation as $\rho_{\rm rad} = \rho_\gamma + \rho_\nu + \rho_{\rm DR}$, where $\rho_\nu$ accounts for the SM neutrino contribution and $\rho_{\rm DR}$ is the DR energy density, $\Delta N_{\rm eff}$ parametrizes the DR component.
Proposed experiments such as CMB-HD, with a projected $2\sigma$ sensitivity of $\Delta N_{\rm eff}\approx 0.027$, may be able to probe the DR shifts predicted in this work.

In this work, DR receives two independent contributions. The first is gravitons emitted via Hawking radiation, whose Kerr enhancement, redshift suppression, and entropy dilution were discussed in sec.~\ref{sec:kerr_pbh}.
The second is GWs radiated by the superradiant cloud as it dissipates, as described in sec.~\ref{sec:superradiance}. Since both channels produce free-streaming radiation that does not thermalize with the SM plasma, we track them independently and write
\begin{equation}
    \Delta N_{\rm eff} = \Delta N_{\rm eff}^{\rm HR} + \Delta N_{\rm eff}^{\rm SR}\,.
    \label{eq:Delta_Neff_split}
\end{equation}

After PBH evaporation is complete, entropy injection from Hawking radiation ends.\footnote{Since the BSM boson decays into SM species (see sec.~\ref{sec:superradiance}), entropy injection from BSM boson decays can occur during PBH evaporation and persist afterwards. In this scenario, $\Delta N_{\rm eff}$ should be evaluated after both PBH evaporation and BSM boson decay are complete, rather than immediately after PBH evaporation.}
The DR energy density, being fully decoupled, redshifts as $\rho_{\rm DR}\propto a^{-4}$. For the SM plasma, the comoving entropy is now conserved, $\mathcal{S}=\mathrm{const}$, and from eq.~\eqref{eq:rho_SM_scaling} $\rho_{\rm SM}$ redshifts as $g_*(T)\,g_{*S}(T)^{-4/3}\,a^{-4}$.
The ratio $\rho_{\rm DR}/\rho_{\rm SM}$ therefore continues to evolve as particle species become non-relativistic and decouple from the thermal bath.
To evaluate $\Delta N_{\rm eff}$, we take this ratio at the evaporation epoch and correct for the change in $g_*$ and $g_{*S}$ between the evaporation temperature $T_{\rm ev}$ and the matter--radiation equality temperature $T_{\rm eq}$:
\begin{equation}
    \Delta N_{\rm eff} = \left[\frac{8}{7}\left(\frac{4}{11}\right)^{\!-4/3}\! + N_{\rm eff}^{\rm SM}\right]
    \frac{\rho_{\rm GW}(T_{\rm ev})}{\rho_{\rm SM}(T_{\rm ev})}\;
    \mathcal{R}(T_{\rm ev},T_{\rm eq})\,,
    \label{eq:Delta_Neff_extract}
\end{equation}
with the degrees-of-freedom correction factor
\begin{equation}
    \mathcal{R}(T_{\rm ev},T_{\rm eq}) \equiv
    \frac{g_*(T_{\rm ev})}{g_*(T_{\rm eq})}
    \left(\frac{g_{*S}(T_{\rm ev})}{g_{*S}(T_{\rm eq})}\right)^{-4/3}\,,
    \label{eq:dof_correction}
\end{equation}
where $\rho_{\rm GW} = \rho_{\rm GW}^{\rm HR} + \sum_i \rho_{{\rm GW},i}^{\rm SR}$ is the total GW energy density from both channels.

It is worth noting that an additional source of DR arises when PBHs temporarily dominate the energy budget before fully evaporating.
The Poisson-distributed spatial number density of PBHs seeds isocurvature density perturbations, which induce a stochastic GW background during the subsequent radiation-dominated era~\cite{Domenech:2020ssp,Papanikolaou:2020qtd,Papanikolaou:2022chm,Domenech:2021wkk}.
This contribution depends mostly on $M_{\rm ini}$ and $\beta$ and can therefore be estimated analytically without modifying the evolution system established here.
The peak spectral density fraction of the induced GWs, evaluated in the post-evaporation era, is~\cite{Domenech:2020ssp}
\begin{equation}
    \Omega_{\rm GW}^{\rm ind} \approx 10^{30}\,\beta^{16/3}
    \left(\frac{\gamma}{0.2}\right)^{8/3}
    \left(\frac{g_H(T_{H, \mathrm{ini}})}{108}\right)^{-17/9}
    \left(\frac{M_{\rm ini}}{10^4\,{\rm g}}\right)^{34/9},
    \label{eq:OmegaGW_ind}
\end{equation}
where $g_H(T_{H, \mathrm{ini}})$ is the spin-weighted number of degrees of freedom contributing to Hawking radiation at PBH formation.
Substituting into eq.~\eqref{eq:Delta_Neff_extract} gives
\begin{equation}
    \Delta N_{\rm eff}^{\rm ind} = \left[\frac{8}{7}\left(\frac{4}{11}\right)^{\!-4/3}\! + N_{\rm eff}^{\rm SM}\right]
    \Omega_{\rm GW}^{\rm ind}\;
    \mathcal{R}(T_{\rm ev},T_{\rm eq})\,.
    \label{eq:Delta_Neff_ind}
\end{equation}
For a benchmark estimate, we adopt $\gamma = 0.2$ and $g_H(T_{H, \mathrm{ini}}) = 108$, so that eq.~\eqref{eq:OmegaGW_ind} reduces to
\begin{equation}
    \left.\Omega_{\mathrm{GW}}^{\mathrm{ind}}\right|_{\text {benchmark }} \approx 9.2 \times 10^{-20}\left(\frac{\beta}{\beta_c}\right)^{16 / 3}\left(\frac{M_{\mathrm{ini}}}{10^4 \mathrm{~g}}\right)^{-14 / 9}.
    \label{eq:OmegaGW_ind_benchmark}
\end{equation}
The corresponding $\Delta N_{\rm eff}$ contribution is then
\begin{equation}
    \left. \Delta N_{\rm eff}^{\rm ind} \right|_{\text {benchmark }} \approx 6.9 \times 10^{-19}\left(\frac{\beta}{\beta_c}\right)^{16 / 3}\left(\frac{M_{\mathrm{ini}}}{10^4 \mathrm{~g}}\right)^{-14 / 9} \mathcal{R}(T_{\rm ev},T_{\rm eq}).
    \label{eq:Delta_Neff_ind_benchmark}
\end{equation}
The correction factor ranges from $\mathcal{R}(T_{\rm ev},T_{\rm eq}) \approx 0.39$ for $M_{\rm ini} \lesssim 10^5~\text{g}$ to $\mathcal{R}(T_{\rm ev},T_{\rm eq}) \approx 0.83$ for $M_{\rm ini} \gtrsim 10^8~\text{g}$.
The steep dependence on $\beta$ can make $\Delta N_{\rm eff}^{\rm ind}$ non-negligible. For instance, $\Delta N_{\rm eff}^{\rm ind} \approx 0.02$ for $\beta = 100\,\beta_c$ and $M_{\rm ini} = 1~\text{g}$.
However, $\Delta N_{\rm eff}^{\rm ind}$ drops rapidly with increasing $M_{\rm ini}$.
In this work we therefore restrict attention to $M_{\rm ini} = 10^2$--$10^9~\text{g}$ with $\beta = 100\,\beta_c$, a parameter range in which the induced-GW contribution is negligible while GWs from Hawking radiation and superradiance can be significant.

\subsection{Comoving entropy and SM heating}\label{sec:entropy}

When $g_*(T)$ and $g_{*S}(T)$ vary, differentiating eq.~\eqref{eq:rho_SM_scaling} gives
\begin{equation}
    \begin{aligned}
        \dot{\rho}_{\rm SM}
         & = \rho_{\rm SM}\!\left(\frac{{\rm d}}{{\rm d}t}\ln\frac{g_*}{g_{*S}^{4/3}}
        + \frac{4}{3}\,\frac{\dot{\mathcal{S}}}{\mathcal{S}} - 4H\right)              \\
         & = -4H\rho_{\rm SM} + Q_{\rm inj}
        + \rho_{\rm SM}\,\frac{{\rm d}}{{\rm d}t}\ln\!\left(\frac{g_*}{g_{*S}^{4/3}}\right),
    \end{aligned}
    \label{eq:rhoSM_corrected}
\end{equation}
where $Q_{\rm inj}\equiv\tfrac{4}{3}(\dot{\mathcal{S}}/\mathcal{S})\,\rho_{\rm SM}$ is the total energy density injection rate into the SM plasma, as follows from the first law ${\rm d}(\rho\,a^3)+p\,{\rm d}(a^3)=T\,{\rm d}\mathcal{S}$ with the radiation enthalpy relation $Ts=(4/3)\rho$.
The last term couples $\dot{\rho}_{\rm SM}$ to $\mathrm{d}g_*/\mathrm{d}T$ and $\mathrm{d}g_{*S}/\mathrm{d}T$, which change abruptly whenever a particle species decouples---most notably at the QCD phase transition---rendering eq.~\eqref{eq:rhoSM_corrected} implicit in $T$ and numerically stiff.

The resolution is to instead evolve the comoving entropy $\mathcal{S}$, which is exactly conserved in the source-free limit regardless of changes in $g_*$ and $g_{*S}$.
This conservation is precisely the origin of the correction factor $\mathcal{R}$ in eq.~\eqref{eq:Delta_Neff_extract}.
Admitting a direct generalization to the sourced case, inverting the definition of $Q_{\rm inj}$ gives $\dot{\mathcal{S}} = \tfrac{3}{4}(Q_{\rm inj}/\rho_{\rm SM})\,\mathcal{S}$.
Substituting $\rho_{\rm SM} = \tfrac{3}{4}(g_*/g_{*S})\,sT$ with $s = \mathcal{S}/a^3$ yields the explicit evolution equation
\begin{equation}
    \frac{{\rm d}\mathcal{S}}{{\rm d}t} = \frac{g_{*S}(T)}{g_*(T)}\,\frac{Q_{\rm inj}}{T}\,a^3\,.
    \label{eq:dSdt}
\end{equation}
When $Q_{\rm inj} = 0$, $\dot{\mathcal{S}} = 0$ and the numerical evolution automatically reproduces the analytic result of eqs.~\eqref{eq:Delta_Neff_extract}--\eqref{eq:dof_correction}.
At each time step, the SM temperature is recovered by inverting $s = (2\pi^2/45)\,g_{*S}(T)\,T^3$ at the current value $s = \mathcal{S}/a^3$, and the SM energy density follows as $\rho_{\rm SM} = (3/4)(g_*/g_{*S})\,sT$.

Two channels contribute to $Q_{\rm inj}$. The first is direct Hawking radiation of SM particles,
\begin{equation}
    Q_{\rm inj}^{\rm HR} = \frac{f_{\rm SM}(M,a_\ast)}{M^2}\,n_{\rm PBH}\,,
    \label{eq:Q_HR}
\end{equation}
where $f_{\rm SM}$ is the SM contribution to the Page function.
The second is the decay of BSM bosons---produced by both Hawking radiation and superradiance---back into SM species,
\begin{equation}
    Q_{\rm inj}^{\rm decay} = \Gamma^{\rm decay}_{\rm BSM}\,\mu\!\left(N_{\rm BSM}^{\rm HR} + \sum_i N_i^{\rm SR}\right) n_{\rm PBH}\,,
    \label{eq:Q_dec}
\end{equation}
where $N_{\rm BSM}^{\rm HR}$ is the number of Hawking-produced BSM bosons per BH.\footnote{For the light PBHs of interest, $\alpha\lesssim\mathcal{O}(1)$ implies $\mu \gtrsim \mathcal{O}(T_H)$, so the BSM boson is heavy relative to the Hawking temperature.
    Its kinetic energy upon emission is therefore subdominant compared to the rest mass, and the total energy of the Hawking-produced BSM population is well approximated by $\mu\,N_{\rm BSM}^{\rm HR}$.}
The total injection rate is then
\begin{equation}
    Q_{\rm inj} = Q_{\rm inj}^{\rm HR} + Q_{\rm inj}^{\rm decay}\,.
    \label{eq:Q_inj}
\end{equation}

Together, eqs.~\eqref{eq:dSdt} and~\eqref{eq:Q_inj} determine the SM thermal history throughout PBH evaporation and will be incorporated into the full evolution system in sec.~\ref{sec:evolution_eqs}.

\subsection{Complete evolution system}\label{sec:evolution_eqs}

The single-BH dynamics developed in Secs.~\ref{sec:kerr_pbh} and~\ref{sec:superradiance} are now embedded in an expanding universe.
The physical PBH number density is $n_{\rm PBH} = \mathcal{N}_{\rm PBH}/a^3$, where the comoving number density $\mathcal{N}_{\rm PBH} \equiv n_{\rm PBH}\,a^3$ is conserved.
We also define the comoving energy densities $\varrho \equiv \rho\,a^3$ for the cosmological components. Therefore, the full coupled system reads
\begin{align}
    \frac{{\rm d}M}{{\rm d}t}                    & = -\frac{f(M,a_\ast)}{M^2} - \sum_i \mu\,\Gamma^{\rm SR}_i\,N^\mathrm{SR}_i\,,\label{eq:dMdt_full}                                      \\
    \frac{{\rm d}J}{{\rm d}t}                    & = -\frac{a_\ast\,g(M,a_\ast)}{M} - \sum_i m_i\,\Gamma^{\rm SR}_i\,N^\mathrm{SR}_i\,,\label{eq:dJdt_full}                                \\
    \frac{{\rm d}N^\mathrm{SR}_i}{{\rm d}t}      & = \Gamma^{\rm SR}_i\,N^\mathrm{SR}_i - \frac{P^{\rm GW}_i}{\mu} - \Gamma^{\rm decay}_{\rm BSM}\,N^\mathrm{SR}_i\,,\label{eq:dNdt_recap} \\
    \frac{{\rm d}N_{\rm BSM}^{\rm HR}}{{\rm d}t} & = \Gamma^{\rm HR}_{\rm BSM} - \Gamma^{\rm decay}_{\rm BSM}\,N_{\rm BSM}^{\rm HR}\,,\label{eq:dNBSM_dt}                                  \\
    \frac{{\rm d}\mathcal{S}}{{\rm d}t}          & = \frac{g_{*S}\left(T\right)}{g_*\left(T\right)}\,\frac{Q_{\rm inj}}{T}\,a^3\,,\label{eq:dSdt_recap}                                    \\
    \frac{{\rm d}\varrho_{\rm GW}^{\rm HR}}{{\rm d}t}
                                                 & = \frac{f_{\rm GW}(M,a_\ast)}{M^2}\,n_{\rm PBH}\,a^3 - H\,\varrho_{\rm GW}^{\rm HR}\,,\label{eq:drhoGW_HR}                              \\
    \frac{{\rm d}\varrho_{{\rm GW},i}^{\rm SR}}{{\rm d}t}
                                                 & = P_i^{\rm GW}\,n_{\rm PBH}\,a^3 - H\,\varrho_{{\rm GW},i}^{\rm SR}\,,\label{eq:drhoGW_SR}                                              \\
    \frac{{\rm d}a}{{\rm d}t}                    & = a\,H\,,\label{eq:dadt_Friedmann}
\end{align}
with the Hubble rate
\begin{equation}
    H^2 = \frac{8\pi}{3}\,\rho_{\rm total}\,,
    \label{eq:Friedmann}
\end{equation}
and the total energy density
\begin{equation}
    \rho_{\rm total} = \rho_{\rm SM} + \rho_{\rm PBH}
    + \rho_{\rm BSM} + \rho_{\rm GW}\,.
    \label{eq:rho_total}
\end{equation}

Eqs.~\eqref{eq:dMdt_full}--\eqref{eq:dNBSM_dt} describe the single-BH dynamics. The BH loses mass and angular momentum through Hawking radiation and superradiance, producing BSM bosons via both channels.
These bosons subsequently decay into SM species at rate $\Gamma^{\rm decay}_{\rm BSM}$, injecting energy into the plasma through $Q_{\rm inj}$ [eqs.~\eqref{eq:Q_HR}--\eqref{eq:Q_inj}], which sources the comoving entropy equation [eq.~\eqref{eq:dSdt_recap}] derived in sec.~\ref{sec:entropy}.

The single-BH rates are promoted to cosmological source terms through the PBH number density $n_{\rm PBH} = \mathcal{N}_{\rm PBH}/a^3$.
In eq.~\eqref{eq:drhoGW_HR}, $f_{\rm GW}(M, a_\ast)$ is the graviton contribution to the Page function $f(M, a_\ast)$.
For the GW energy densities, the comoving variable $\varrho \equiv \rho\,a^3$ absorbs the $a^{-3}$ dilution; the source terms in eqs.~\eqref{eq:drhoGW_HR}--\eqref{eq:drhoGW_SR} represent GW energy injection per comoving volume, while the $-H\varrho$ terms account for the additional $a^{-1}$ redshift of massless radiation.
Tracking $\varrho_{\rm GW}$ in real time is essential for correctly capturing the differential redshift discussed in sec.~\ref{sec:kerr_pbh}.

In $\rho_{\rm total}$, $\rho_{\rm SM}$ is reconstructed from $\mathcal{S}$ and $a$ as in sec.~\ref{sec:entropy}, $\rho_{\rm PBH} = M\,n_{\rm PBH}$, while $\rho_{\rm BSM}$ and $\rho_{\rm GW}$ each collect contributions from both the Hawking and superradiant channels.

The above system has five free parameters:\label{sec:numerics} the initial PBH mass $M_{\rm ini}$, the initial dimensionless spin $a_{*,\rm ini}$, the initial PBH abundance $\beta$, the gravitational coupling $\alpha_{\rm ini} \equiv M_{\rm ini}\mu$, and the BSM boson spin $s$.
We scan $M_{\rm ini}$ over $\sim 0.1$--$10^9\,{\rm g}$, the range in which PBHs evaporate completely before BBN.
The spin $a_{*,\rm ini}$ ranges from the near-Schwarzschild limit ($a_{*,\rm ini} \approx 0$) to near-extremal Kerr ($a_{*,\rm ini} \lesssim 1$).
The abundance $\beta$ covers both the radiation-dominated regime ($\beta < \beta_c$) and the regime where PBHs eventually overtake radiation and temporarily dominate the energy budget ($\beta > \beta_c$).
In the latter regime, the evaporation products reset the radiation content, and $\Delta N_{\rm eff}$ becomes insensitive to the precise value of $\beta$.
The coupling $\alpha_{\rm ini}$ sets the superradiance timescale.
We consider both scalar ($s = 0$) and vector ($s = 1$) BSM bosons, which differ in their mode evolutions and cloud GW emissions.

Given $M_{\rm ini}$, the PBH formation temperature $T_{\rm ini}$ is fixed by the horizon mass relation~\eqref{eq:Mini}.
The initial comoving PBH number density $\mathcal{N}_{\rm PBH}$ and the SM radiation energy density are then determined by $\beta$ and $T_{\rm ini}$.
The scale factor is normalized to $a_{\rm ini} = 1$ at formation.
Each superradiant mode is seeded with a small initial occupation number; the exact value is irrelevant because the subsequent exponential growth quickly erases any dependence on the seed.
The coupled ODE system~\eqref{eq:dMdt_full}--\eqref{eq:dadt_Friedmann} is solved with a stiff integrator.
The Hawking radiation functions $f(M, a_\ast)$, $g(M, a_\ast)$, and $f_{\rm GW}(M, a_\ast)$ are evaluated from pre-computed greybody-factor tables provided by \texttt{BlackHawk}~\cite{Arbey:2019mbc, Arbey:2021mbl} and \texttt{FRISBHEE}~\cite{Cheek:2021odj, Cheek:2021cfe, Cheek:2022dbx, Cheek:2022mmy}.
The thermodynamic functions $g_*(T)$ and $g_{*S}(T)$ are constructed by interpolating a composite table that draws on lattice QCD results from Borsanyi et al.~\cite{Borsanyi:2016ksw} for the crossover region $1\,{\rm MeV} < T < 500\,{\rm MeV}$, supplemented by Husdal~\cite{Husdal:2016haj} at higher and lower temperatures (see also ref.~\cite{Wallisch:2018rzj}).

\section{Results}\label{sec:results}

\subsection{Representative evolution}\label{sec:representative}

\begin{figure*}
    \includegraphics[width=0.48\textwidth]{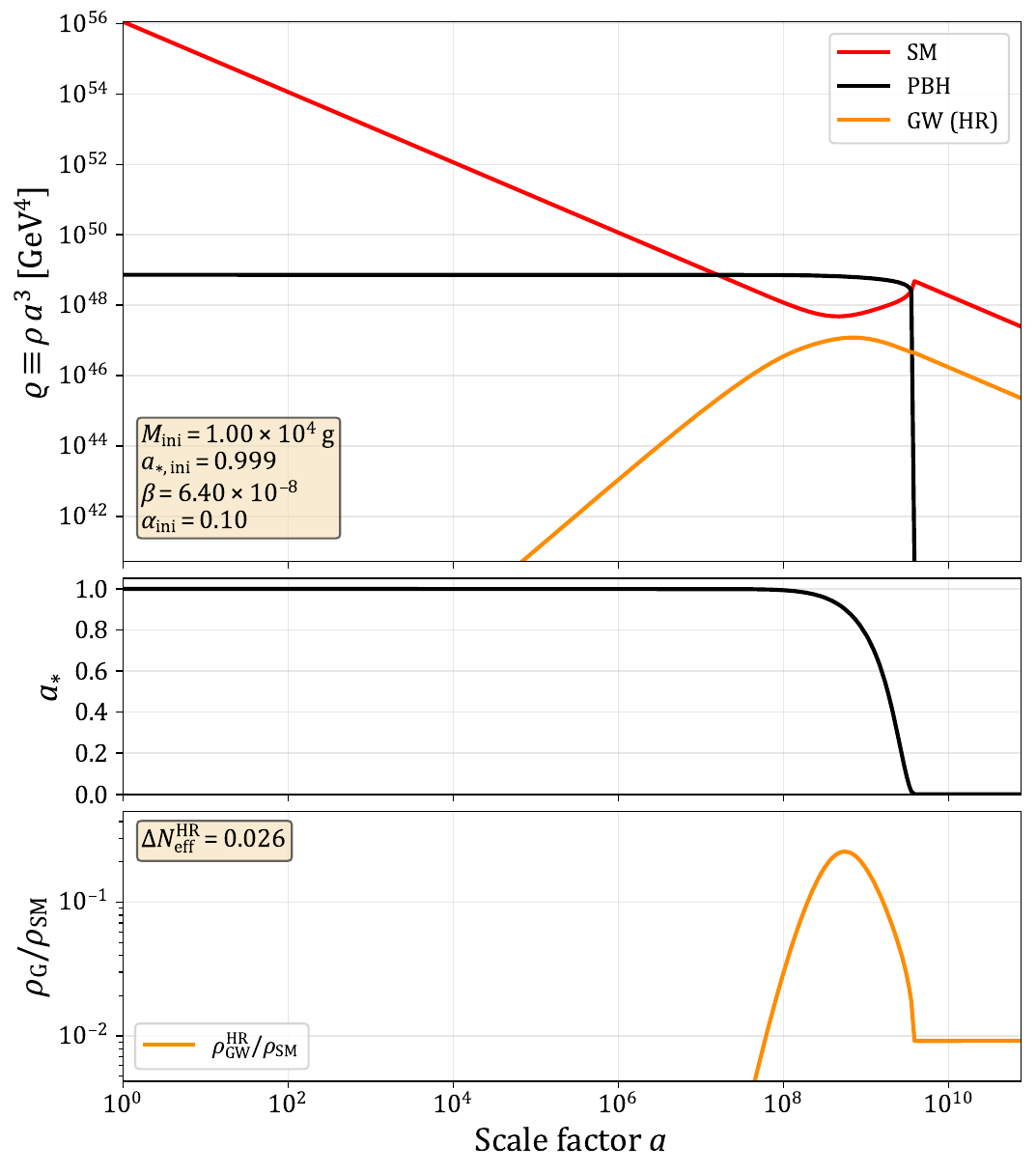}\hfill
    \includegraphics[width=0.48\textwidth]{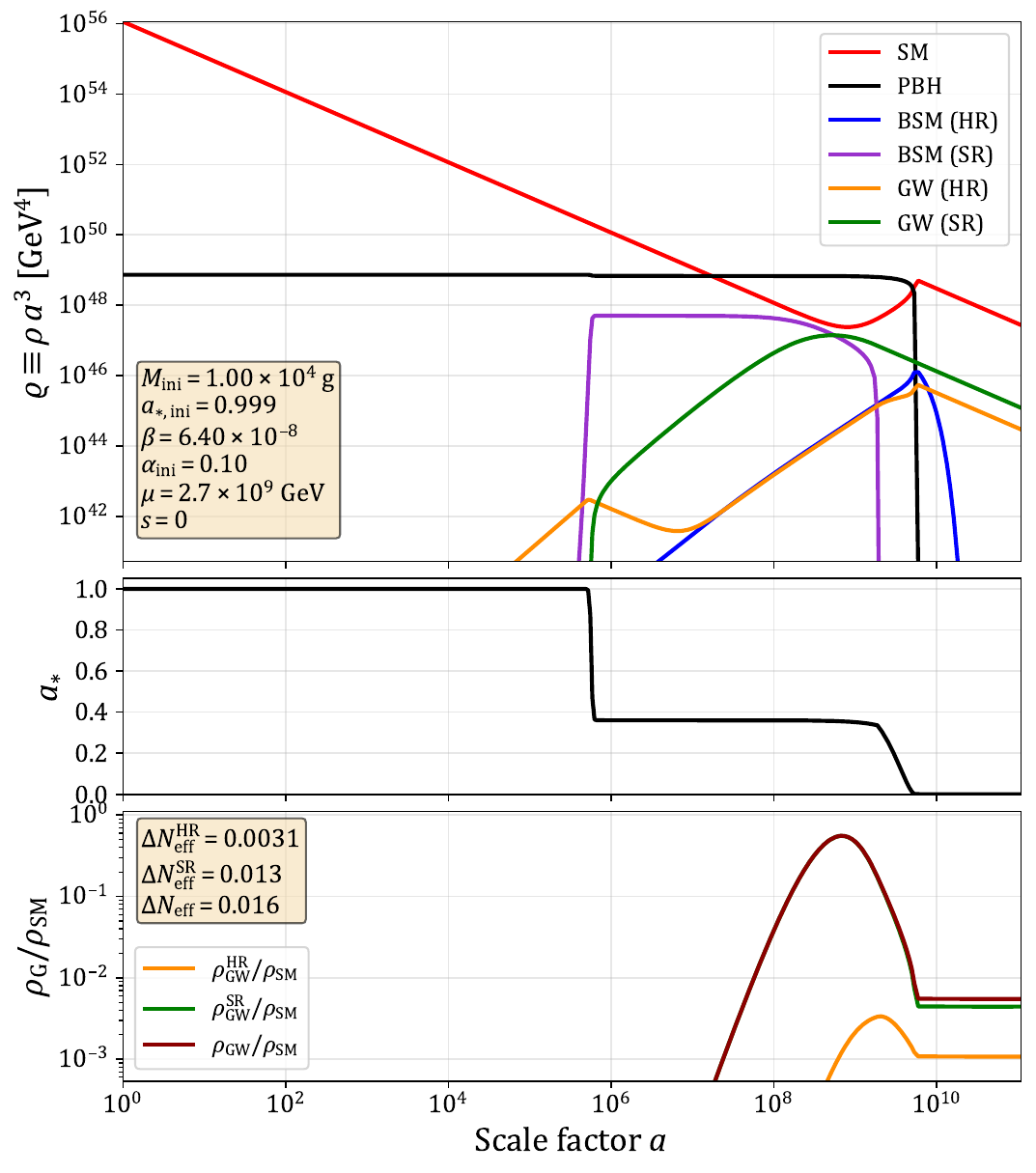}
    \caption{Cosmological evolution of the comoving energy densities $\varrho \equiv \rho\,a^3$ (top), the BH spin parameter $a_\ast$ (middle), and the GW-to-SM energy density ratio $\rho_{\rm GW}/\rho_{\rm SM}$ (bottom) as a function of the scale factor~$a$, for a benchmark with $M_{\rm ini} = 10^4\,{\rm g}$, $a_{*,\rm ini} = 0.999$, $\beta = 6.40 \times 10^{-8}$, $\alpha_{\rm ini} = 0.10$ ($\mu = 2.7 \times 10^{9}\,{\rm GeV}$, scalar $s = 0$).
        \textit{Left:} Hawking-radiation-only baseline. The SM radiation (red) and PBH matter (black) components cross near $a \sim 10^7$, producing a brief PBH-dominated epoch that ends with evaporation near $a \sim 10^{10}$. The spin remains near-extremal throughout most of the evolution, dropping only at the very end of evaporation. The Hawking graviton energy density (GW~HR, orange) peaks near evaporation.
        \textit{Right:} Full calculation including superradiance. The superradiant scalar cloud (BSM~SR, purple) grows rapidly before evaporation, becoming the third-largest energy component. The middle subpanel shows the characteristic staircase spin-down: $a_\ast$ drops sharply from $a_{*,\rm ini}$ to $a_{\ast c}^{|211\rangle}$ as the superradiant mode extracts angular momentum, well before Hawking evaporation sets in. The associated GW emission (GW~SR, green) is one to two orders of magnitude below the cloud energy, since only a fraction of the cloud radiates gravitationally. The accelerated spin-down suppresses the Hawking graviton channel, and the total $\Delta N_{\rm eff}$ is reduced relative to the left panel despite the new superradiant cloud GW contribution.}
    \label{fig:benchmark}
\end{figure*}

To illustrate the interplay between Hawking radiation and superradiance, we select a benchmark parameter point with $M_{\rm ini} = 10^4\,{\rm g}$, $a_{*,\rm ini} = 0.999$, $\beta = 100 \beta_c = 6.40 \times 10^{-8}$, and a scalar ($s = 0$) BSM boson with $\alpha_{\rm ini} = 0.10$ ($\mu = 2.7 \times 10^{9}\,{\rm GeV}$).
The near-extremal spin maximizes the Hawking graviton emission rate, $\alpha_{\rm ini} = 0.10$ ensures efficient superradiant extraction, and $\beta \gg \beta_c$ guarantees a PBH-dominated epoch so that $\Delta N_{\rm eff}$ is independent of the precise $\beta$ value (cf.\ sec.~\ref{sec:numerics}).
Figure~\ref{fig:benchmark} compares the cosmological evolution with Hawking radiation only (left) and with superradiance included (right).

In the left panel, the comoving SM energy density $\varrho_{\rm SM}$ decreases as $a^{-1}$, while the PBH component $\varrho_{\rm PBH}$ remains constant.
The two cross near $a \sim 10^7$, after which PBHs briefly dominate the energy budget until evaporation near $a \sim 10^{10}$.
As discussed in sec.~\ref{sec:kerr_pbh}, the near-extremal initial spin enhances graviton emission during the early stage, which is brief relative to the total BH lifetime ($\tau_a \ll \tau_M$).
The middle subpanel confirms that $a_\ast$ remains near unity throughout most of the evolution, dropping to zero only at the very end of evaporation.
The resulting GW energy density peaks near the depletion of BH spin and subsequently redshifts as radiation.
The lower subpanel shows that $\rho_{\rm GW}^{\rm HR}/\rho_{\rm SM}$ freezes out to give $\Delta N_{\rm eff}^{\rm HR} = 0.026$, nearly coinciding with the projected $2\sigma$ sensitivity of CMB-HD.

The right panel shows the same system with superradiance active.
Well before evaporation, at $a \sim 10^5$, the superradiant instability begins transferring angular momentum and energy into the scalar cloud.
The spin evolution (middle subpanel) shows a sharp drop to $a_{\ast c}^{|211\rangle} \approx 0.384$ [eq.~\eqref{eq:a_crit}], below which superradiant extraction ceases.
The cloud energy density (BSM~SR, purple) rises steeply, becoming the third-largest component after SM radiation and PBHs, then falls rapidly once $a_\ast$ drops below $a_{\ast c}$.
This rapid decline is dominated by superradiant decay: the BH re-absorbs the cloud bosons far faster than GW emission or intrinsic BSM boson decay can drain them.
Since the cloud GW power is proportional to the cloud mass, the GW energy density (GW~SR, green) grows efficiently only after the cloud reaches large occupation numbers.
Superradiance and Hawking radiation both extract angular momentum from the BH [eq.~\eqref{eq:dJdt_full}], but once the cloud occupation number is large the superradiant extraction rate $\sum_i m_i\,\Gamma_i^{\rm SR}\,N_i^{\rm SR}$ dominates over the Hawking contribution $a_\ast\,g(M,a_\ast)/M$, so superradiance captures most of the available spin before Hawking radiation can convert it into gravitons.
As a result, $\Delta N_{\rm eff}^{\rm HR}$ drops to $0.0031$, a reduction by roughly an order of magnitude.
The superradiant GW contribution partially compensates, adding $\Delta N_{\rm eff}^{\rm SR} = 0.013$, but the total $\Delta N_{\rm eff} = 0.016$ remains below the pure Hawking-radiation value.

The net reduction reflects a redistribution of the energy budget by superradiance.
Energy is diverted from the Hawking radiation GW channel into the scalar cloud, where the bulk of it eventually decays back into SM species, reheating the plasma rather than contributing to DR.
Two effects compete: the superradiant cloud provides a new GW source, while superradiance starves the Hawking channel of the BH spin needed for efficient graviton production.
For this benchmark, the suppression dominates.

\begin{figure*}
    \includegraphics[width=0.48\textwidth]{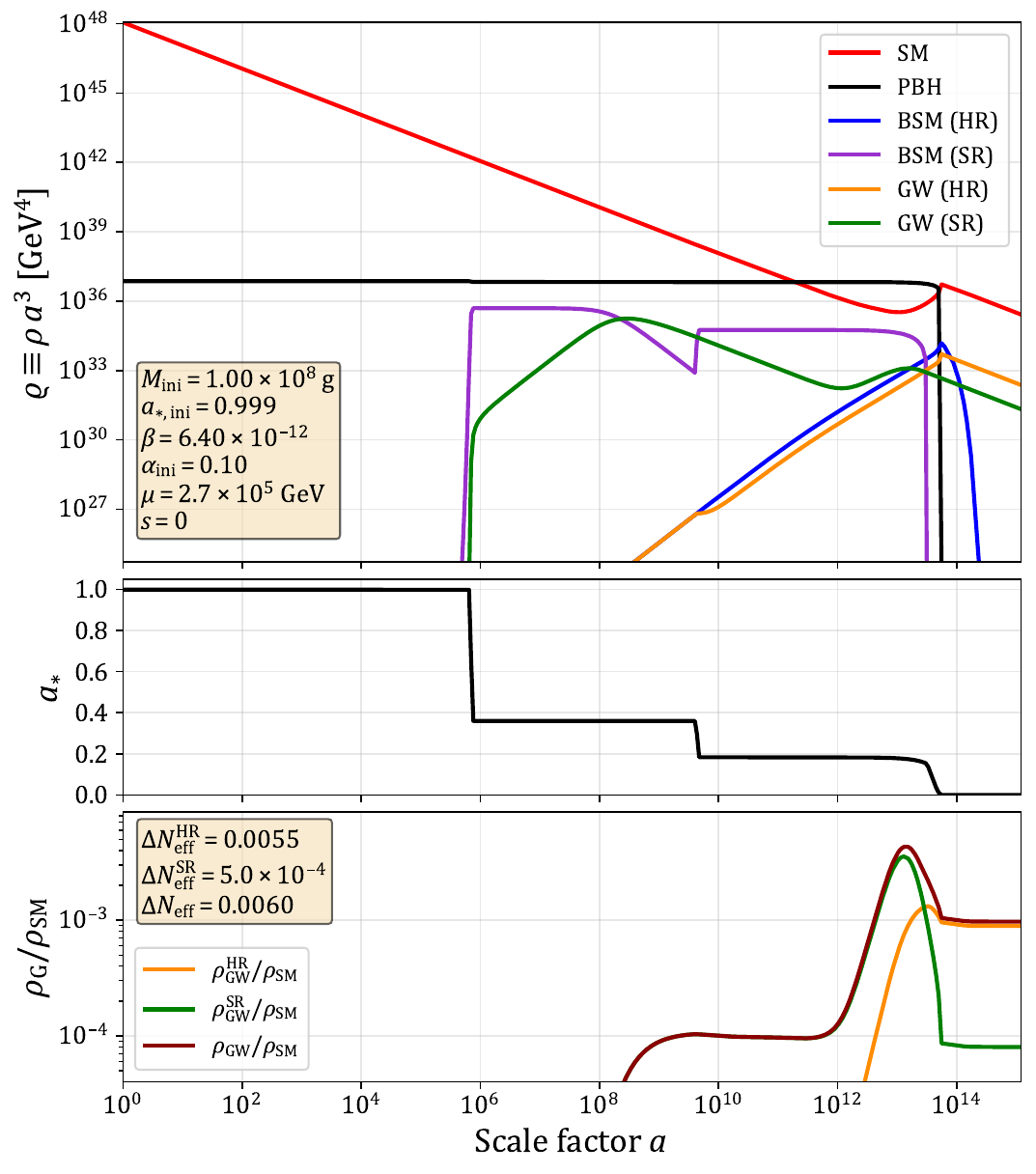}\hfill
    \includegraphics[width=0.48\textwidth]{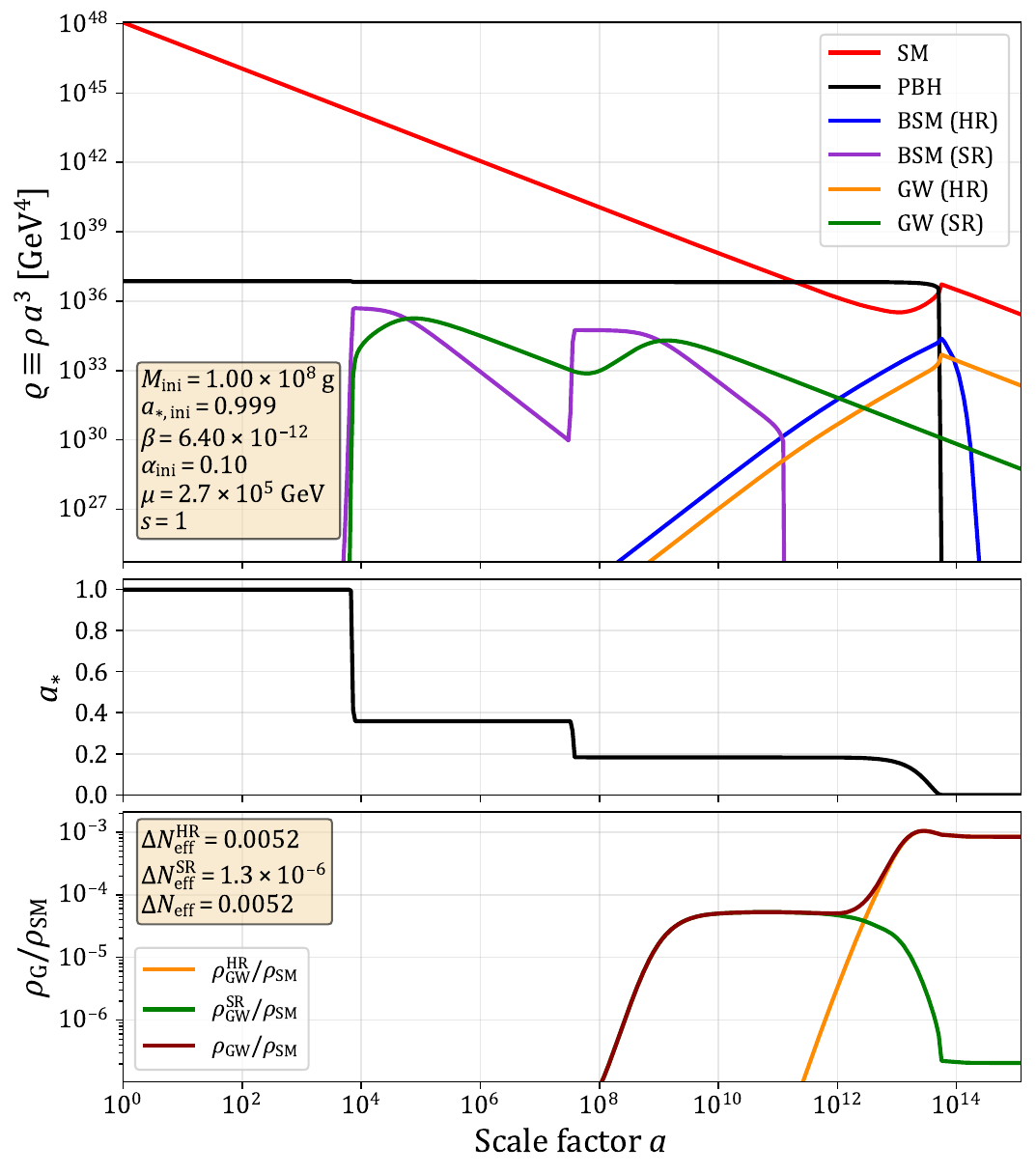}
    \caption{Same format as fig.~\ref{fig:benchmark} but for $M_{\rm ini} = 10^8\,{\rm g}$, $\beta = 6.40 \times 10^{-12}$, $\alpha_{\rm ini} = 0.10$ ($\mu = 2.7 \times 10^{5}\,{\rm GeV}$), with $a_{*,\rm ini} = 0.999$ and superradiance active in both panels.
        \textit{Left:} Scalar boson ($s = 0$). The two-step staircase in $a_\ast$ (middle subpanel) directly traces the sequential mode dynamics: the $|211\rangle$ mode grows first (first bump in BSM~SR near $a \sim 10^7$), spinning down $a_\ast$ to $a_{\ast c}^{|211\rangle}$, and the $|322\rangle$ mode subsequently takes over (second bump near $a \sim 10^{11}$), reducing $a_\ast$ further to $a_{\ast c}^{|322\rangle}$. Each mode produces a corresponding feature in GW~SR (green), but the combined $\Delta N_{\rm eff}^{\rm SR} = 5.0 \times 10^{-4}$ is far below the Hawking channel.
        \textit{Right:} Vector boson ($s = 1$). The dominant $|1011\rangle$ mode initiates superradiance much earlier ($a \sim 10^3$) due to its faster growth rate, as reflected by the earlier spin-down in the middle subpanel. The GWs emitted at this early epoch undergo far greater redshift before freeze-out, reducing the SR contribution to $\Delta N_{\rm eff}^{\rm SR} = 1.3 \times 10^{-6}$.}
    \label{fig:multimode}
\end{figure*}

To resolve the multimode dynamics, we increase the PBH mass to $M_{\rm ini} = 10^8\,{\rm g}$.
Since $\tau_{\rm BH} \propto M^3$, the BH lifetime is $\sim 10^{12}$ times longer than the $10^4\,{\rm g}$ benchmark, giving the secondary superradiant mode time to complete its full growth--saturation--GW-emission cycle before evaporation.
Figure~\ref{fig:multimode} compares a scalar (left) and a vector (right) BSM boson, with all other parameters matching those of fig.~\ref{fig:benchmark}.

In the scalar case (left panel), the two-step staircase in $a_\ast$ (middle subpanel) directly traces the sequential mode dynamics.
The $|211\rangle$ mode grows first, extracting angular momentum until $a_\ast$ drops to $a_{\ast c}^{|211\rangle}$ and its growth stalls.
The $|322\rangle$ mode, whose critical spin $a_{\ast c}^{|322\rangle} < a_{\ast c}^{|211\rangle}$, then takes over and extracts further angular momentum, producing the second step in the spin history.
Each mode produces a distinct bump in the cloud energy density (BSM~SR, purple) and the GW emission (GW~SR, green).
However, the first mode's GWs are emitted around $a \sim 10^7$ and redshift by a factor of $\sim 10^7$ before freeze-out near $a \sim 10^{14}$, while the second mode's GW power is suppressed by additional powers of $\alpha$.
The combined superradiant contribution is $\Delta N_{\rm eff}^{\rm SR} = 5.0 \times 10^{-4}$, an order of magnitude below $\Delta N_{\rm eff}^{\rm HR} = 0.0055$.

The vector case (right panel) illustrates a further suppression.
The dominant vector mode $|1011\rangle$ grows faster than the scalar $|211\rangle$, so superradiance initiates at $a \sim 10^3$---orders of magnitude earlier than in the scalar case, as the correspondingly earlier spin-down in the middle subpanel confirms.
This earlier onset suppresses $\Delta N_{\rm eff}^{\rm SR}$: GWs emitted at $a \sim 10^3$ redshift by $\sim 10^{11}$ before freeze-out, reducing $\Delta N_{\rm eff}^{\rm SR}$ to $1.3 \times 10^{-6}$.
The total $\Delta N_{\rm eff} = 0.0052$ is dominated entirely by Hawking radiation.

Comparing Figs.~\ref{fig:benchmark} and~\ref{fig:multimode} reveals a general trend: faster superradiant growth does not translate into a larger GW contribution to DR.
In the $10^4\,{\rm g}$ benchmark, superradiance and evaporation occur on comparable timescales, and the superradiant cloud GWs undergo relatively modest redshift, yielding $\Delta N_{\rm eff}^{\rm SR} = 0.013$.
For $10^8\,{\rm g}$ PBHs, the superradiant episode completes well before evaporation, and the intervening expansion dilutes the GW energy density by orders of magnitude.
The larger the ratio $\tau_{\rm BH}/\tau_{\rm SR}$, the more the superradiant cloud GWs are diluted, making the penalty most severe for vector bosons whose dominant mode has the fastest growth rate.

\subsection{Parameter dependence of \texorpdfstring{$\Delta N_{\rm eff}$}{Delta Neff}}\label{sec:scan}

\begin{figure*}
    \includegraphics[width=0.48\textwidth]{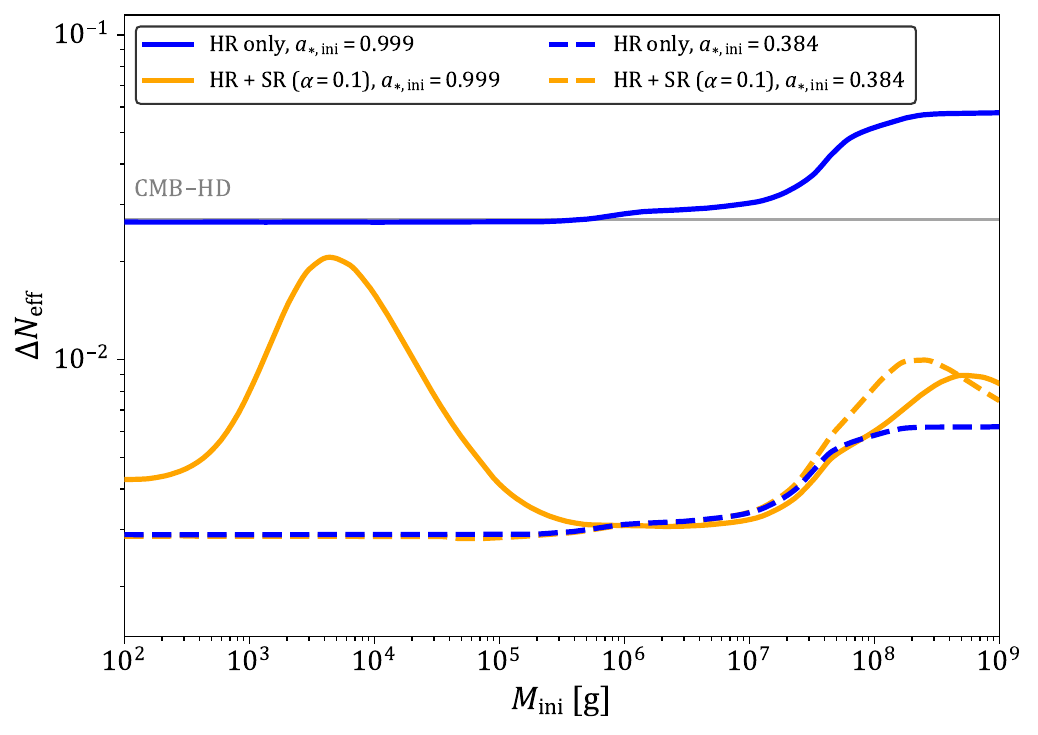}\hfill
    \includegraphics[width=0.48\textwidth]{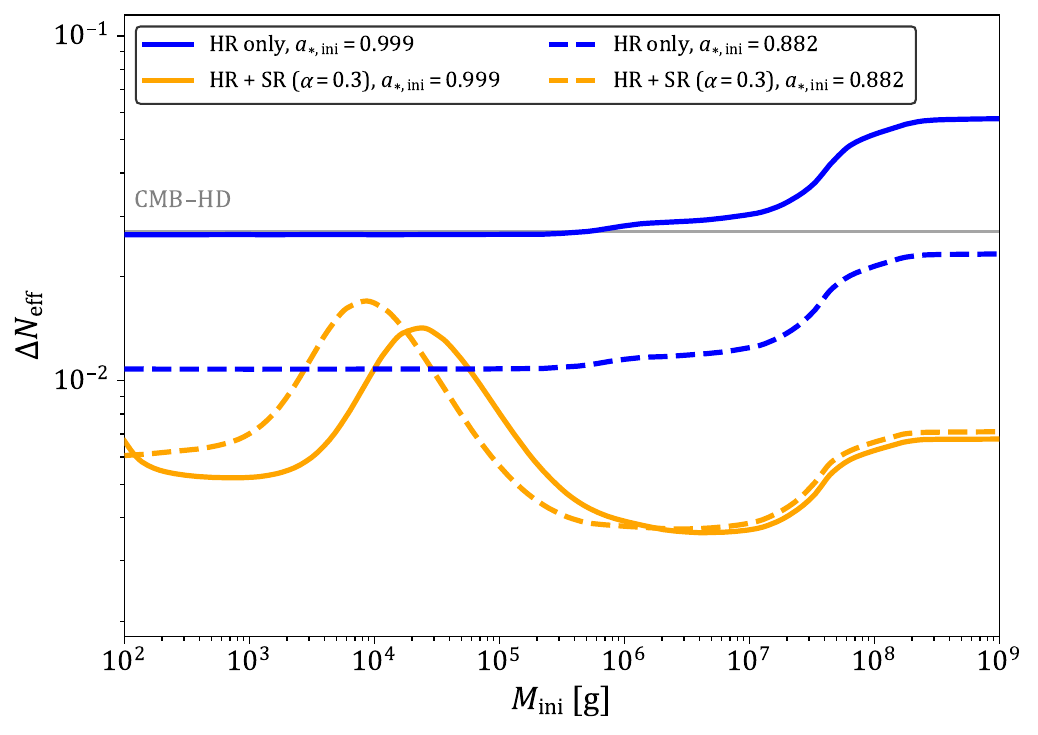}
    \caption{$\Delta N_{\rm eff}$ as a function of $M_{\rm ini}$ for fixed gravitational coupling $\alpha_{\rm ini} = 0.1$ (\textit{left}) and $\alpha_{\rm ini} = 0.3$ (\textit{right}), with $\beta > \beta_c$ so that a PBH-dominated epoch precedes evaporation.
        Blue curves show the Hawking-radiation-only baselines; orange curves include superradiance.
        Solid and dashed lines correspond to the two initial spins indicated in each legend.
        In each panel, the lower spin coincides with the critical spin $a_{\ast c}$ of the $|211\rangle$ mode [eq.~\eqref{eq:a_crit}]: $a_{*,\rm ini} = 0.384$ for $\alpha = 0.1$ (\textit{left}) and $a_{*,\rm ini} = 0.882$ for $\alpha = 0.3$ (\textit{right}), at which the superradiant growth rate vanishes.
        The gray line marks the projected $2\sigma$ sensitivity of CMB-HD ($\Delta N_{\rm eff} \approx 0.027$).}
    \label{fig:scan}
\end{figure*}

Figure~\ref{fig:scan} scans $\Delta N_{\rm eff}$ across the PBH mass range for two representative values of the gravitational coupling: $\alpha_{\rm ini} = 0.1$ (left) and $\alpha_{\rm ini} = 0.3$ (right).
Since $\alpha_{\rm ini} = M_{\rm ini}\mu$, each point along the horizontal axis corresponds to a different BSM boson mass $\mu = \alpha_{\rm ini}/M_{\rm ini}$.
All scans fix $\beta$ above $\beta_c$.

The Hawking-radiation-only baselines (blue) provide the reference against which superradiant effects are measured.
For near-extremal spin $a_{*,\rm ini} = 0.999$ (solid), $\Delta N_{\rm eff}$ lies just below the CMB-HD threshold across $M_{\rm ini} \lesssim 10^5\,{\rm g}$, and rises to $\sim 0.06$ at $10^9\,{\rm g}$.
The rise at high masses reflects the correction factor $\mathcal{R}(T_{\rm ev},T_{\rm eq})$ [eq.~\eqref{eq:dof_correction}]: heavier PBHs have lower Hawking temperatures and evaporate when fewer SM species are relativistic, reducing $g_*(T_{\rm ev})$ and increasing $\mathcal{R}$.
For $a_{*,\rm ini} = a_{\ast c}^{|211\rangle} = 0.384$ (dashed, left panel), chosen so that the dominant superradiant mode is marginally inactive, the low spin strongly suppresses the Hawking graviton emission, giving $\Delta N_{\rm eff} \sim 3 \times 10^{-3}$, an order of magnitude below the near-extremal case.

In the left panel ($\alpha_{\rm ini} = 0.1$), the near-extremal curve with superradiance (orange solid) shows three distinct regimes.
At low masses ($M_{\rm ini} \sim 10^3$--$10^4\,{\rm g}$), a broad bump peaks at $\Delta N_{\rm eff} \approx 0.021$ near $M_{\rm ini} \sim 4 \times 10^3\,{\rm g}$, still below CMB-HD but partially compensating the superradiant suppression.
Here $\tau_{\rm SR}$ and $\tau_{\rm BH}$ are comparable, so the cloud-emitted GWs undergo only moderate redshift, as illustrated by the $10^4\,{\rm g}$ benchmark in fig.~\ref{fig:benchmark}.
At intermediate masses ($10^4$--$10^6\,{\rm g}$), $\Delta N_{\rm eff}$ drops to $\sim 3 \times 10^{-3}$---an order of magnitude below the Hawking-only baseline.
Superradiance completes well before evaporation in this range: the cloud GWs are heavily diluted by subsequent expansion, while the early spin extraction starves the Hawking graviton channel.
At high masses ($\gtrsim 10^8\,{\rm g}$), the curve recovers to $\sim 8 \times 10^{-3}$.
The curve with superradiance at $a_{*,\rm ini} = a_{\ast c}^{|211\rangle}$ (orange dashed) is nearly indistinguishable from its Hawking-only counterpart for $M_{\rm ini} \lesssim 10^6\,{\rm g}$.
At higher masses, the longer BH lifetime allows superradiance to operate, and the cloud GWs push $\Delta N_{\rm eff}$ above the Hawking-only level to a peak of $\sim 0.010$ near $M_{\rm ini} \sim 2.5 \times 10^8\,{\rm g}$---a net enhancement, since the low initial spin leaves little Hawking graviton flux to suppress.

The right panel ($\alpha_{\rm ini} = 0.3$) shows qualitatively similar features with important quantitative shifts.
For $a_{*,\rm ini} = 0.999$ (orange solid), the low-mass bump moves to $M_{\rm ini} \sim 2.4 \times 10^4\,{\rm g}$ and its peak drops to $\sim 0.014$, well below CMB-HD.
A larger $\alpha$ shortens $\tau_{\rm SR}$, so superradiance completes before evaporation even for lighter PBHs, extending the suppression region to lower masses and shifting the bump to higher masses where $\tau_{\rm SR} \sim \tau_{\rm BH}$ still holds.

The second spin in the right panel, $a_{*,\rm ini} = a_{\ast c}^{|211\rangle} = 0.882$ (higher than the left-panel value of $0.384$ because $a_{\ast c}$ grows with $\alpha$), sits at the superradiant threshold.
The $|211\rangle$ mode is therefore inactive, but the subdominant $|322\rangle$ mode, whose $a_{\ast c}^{|322\rangle} < a_{*,\rm ini}$, remains active and drives the superradiant evolution.
The resulting curve with superradiance (orange dashed) departs significantly from its Hawking-only baseline (blue dashed).
At low masses ($M_{\rm ini} \lesssim 10^3\,{\rm g}$), the incomplete superradiance extracts spin and reduces $\Delta N_{\rm eff}$ from $\sim 0.011$ (Hawking-only) to $\sim 0.006$.
A bump peaks at $\Delta N_{\rm eff} \approx 0.017$ near $M_{\rm ini} \sim 8.6 \times 10^3\,{\rm g}$, exceeding the Hawking-only baseline: the superradiant cloud GWs more than compensate the reduced Hawking graviton output.
At high masses the superradiant GW contribution is again diluted by redshift and $\Delta N_{\rm eff}$ falls back below the baseline.

Comparing the two panels, several common trends emerge.
For near-extremal PBHs, superradiance suppresses $\Delta N_{\rm eff}$ to well below the CMB-HD threshold across the full mass range, with partial compensation confined to a narrow window where $\tau_{\rm SR} \sim \tau_{\rm BH}$.
Increasing $\alpha$ sharpens this suppression: the compensating bump narrows, shifts to higher masses, and its peak drops from $0.021$ to $0.014$.
At high masses ($M_{\rm ini} \gtrsim 10^8\,{\rm g}$), all curves with superradiance converge to $\Delta N_{\rm eff} \sim 0.007$--$0.008$ regardless of $\alpha$ or $a_{*,\rm ini}$, because the superradiant GWs are emitted too early and diluted beyond relevance.
The near-extremal Hawking-only baseline sits nearly at the CMB-HD threshold for $M_{\rm ini} \lesssim 10^7\,{\rm g}$ and rises above it at higher masses; superradiance removes this marginal detectability entirely, pushing $\Delta N_{\rm eff}$ below the sensitivity line across the full parameter space explored here.

\subsection{Sensitivity in the \texorpdfstring{$(M_{\rm ini},\,a_{*,\rm ini})$}{(Mini, a*ini)} plane}\label{sec:contours}

\begin{figure*}
    \includegraphics[width=0.48\textwidth]{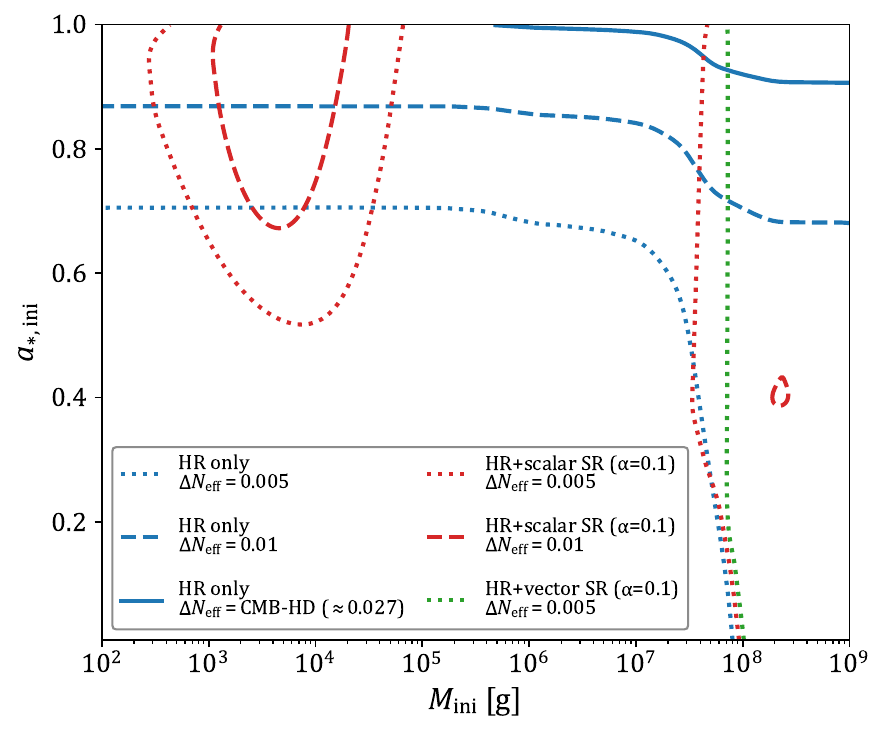}\hfill
    \includegraphics[width=0.48\textwidth]{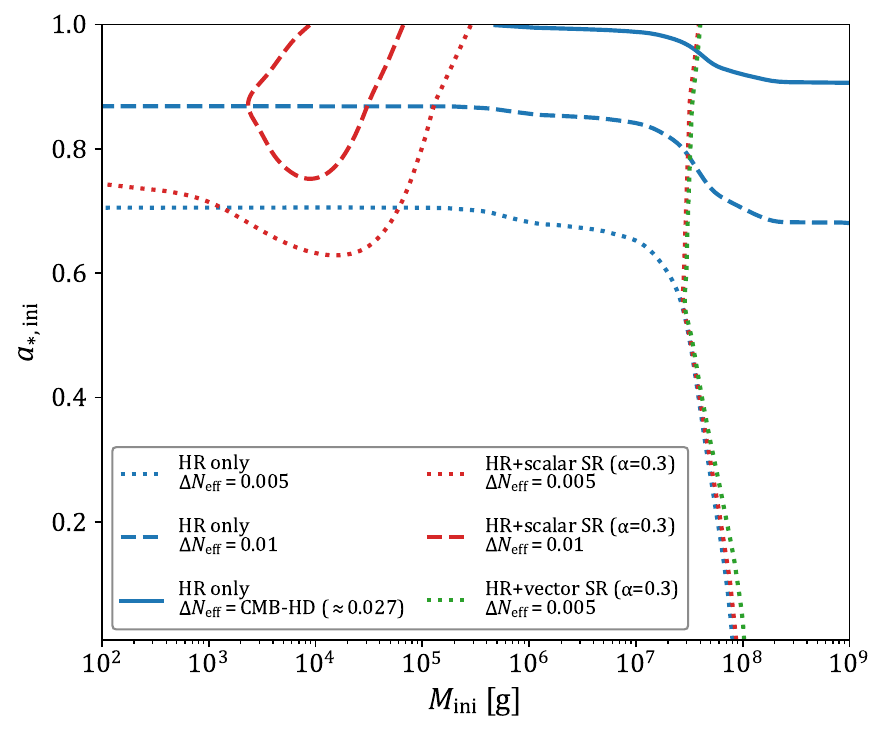}
    \caption{Iso-$\Delta N_{\rm eff}$ contours in the $(M_{\rm ini},\,a_{*,\rm ini})$ plane for $\alpha_{\rm ini} = 0.1$ (\textit{left}) and $\alpha_{\rm ini} = 0.3$ (\textit{right}).
        Blue curves show the Hawking-radiation-only contours at $\Delta N_{\rm eff} = 0.005$ (dotted), $0.01$ (dashed), and $0.027 \approx$ CMB-HD $2\sigma$ sensitivity (solid).
        Red curves show the corresponding contours with scalar ($s=0$) superradiance included, at $\Delta N_{\rm eff} = 0.005$ (dotted) and $0.01$ (dashed).
        The green dashed curve shows $\Delta N_{\rm eff} = 0.005$ for vector ($s=1$) superradiance.
        The region above each contour corresponds to larger $\Delta N_{\rm eff}$.
        No CMB-HD contour exists for either superradiant scenario.}
    \label{fig:contours}
\end{figure*}

The one-dimensional scans of fig.~\ref{fig:scan} fixed $a_{*,\rm ini}$ and varied $M_{\rm ini}$.
Figure~\ref{fig:contours} extends the analysis to the full two-dimensional parameter space by plotting iso-$\Delta N_{\rm eff}$ contours in the $(M_{\rm ini},\,a_{*,\rm ini})$ plane.
As in sec.~\ref{sec:scan}, $\alpha_{\rm ini}$ is held fixed at $0.1$ (left) and $0.3$ (right), and $\beta > \beta_c$ throughout.

The Hawking-only contours (blue) are nearly horizontal for $M_{\rm ini} \lesssim 10^7\,{\rm g}$.
Even at near-extremal spin, $\Delta N_{\rm eff}$ lies just below the CMB-HD threshold in this range.
At higher masses the contours curve downward, because the degree-of-freedom correction $\mathcal{R}$ increases $\Delta N_{\rm eff}$ at fixed spin, so a lower $a_{*,\rm ini}$ suffices to reach the same threshold.
The CMB-HD contour (blue solid) only appears for $M_{\rm ini} \gtrsim 10^7\,{\rm g}$, where the $\mathcal{R}$ correction lifts $\Delta N_{\rm eff}$ above the threshold; lighter PBHs fall just short even at near-extremal spin.

With scalar superradiance included (red), no CMB-HD contour appears in either panel---scalar superradiance pushes $\Delta N_{\rm eff}$ below $0.027$ across the entire $(M_{\rm ini},\,a_{*,\rm ini})$ plane, closing the marginal detectability window that exists in the Hawking-only scenario.

In a narrow mass window where $\tau_{\rm SR} \sim \tau_{\rm BH}$, the superradiant GW bump raises $\Delta N_{\rm eff}$ above the Hawking-only value at the same spin, causing the red contours to dip below their blue counterparts; for $\alpha_{\rm ini} = 0.1$, this produces a small closed island of $\Delta N_{\rm eff} = 0.01$.
This is the two-dimensional counterpart of the bumps identified in fig.~\ref{fig:scan}.

For vector superradiance (green), only the $\Delta N_{\rm eff} = 0.005$ contour can be drawn: the superradiant GW contribution is negligible due to the redshift penalty discussed in sec.~\ref{sec:representative}, and $\Delta N_{\rm eff}$ is dominated entirely by the weakened Hawking graviton channel.

\section{Discussion and conclusions}\label{sec:conclusions}

We have presented a comprehensive calculation of $\Delta N_{\rm eff}$ from Kerr PBHs that simultaneously tracks Hawking radiation and superradiant instability within an expanding cosmological background.
Our framework couples the BH mass and angular momentum evolution, the superradiant mode occupation numbers, the comoving SM entropy, and the GW energy densities from both Hawking radiation and cloud dissipation, consistently accounting for the differential redshift experienced by gravitational radiation emitted at different epochs.
We have examined both scalar ($s=0$) and vector ($s=1$) BSM bosons, each with two superradiant modes, at gravitational couplings $\alpha_{\rm ini}=0.1$ and $0.3$.

The results reveal several robust features.
First, superradiance generically suppresses $\Delta N_{\rm eff}$ relative to the Hawking-only baseline.
By extracting angular momentum before Hawking radiation can convert it into gravitons, superradiance starves the dominant Hawking dark-radiation channel.
The GWs emitted by the superradiant cloud partially compensate this loss, but the net effect remains a reduction.
At our benchmark point (fig.~\ref{fig:benchmark}), $\Delta N_{\rm eff}$ drops from $0.026$ (Hawking-only) to $0.016$ (with superradiance).

Second, faster superradiant growth does not translate into a larger DR contribution.
The dominant vector mode grows more rapidly than its scalar counterpart, but emits its GWs correspondingly earlier.
The additional cosmological redshift renders the vector superradiant GW contribution negligible, while the scalar superradiant GW contribution is significant only in a narrow mass window where $\tau_{\rm SR}\sim\tau_{\rm BH}$.
This redshift penalty is the central mechanism that limits the role of superradiance as a dark-radiation source.

Third, scalar superradiance closes the CMB-HD detectability window for $\Delta N_{\rm eff}$ from Kerr PBHs.
In the Hawking-only scenario, near-extremal PBHs can produce $\Delta N_{\rm eff}$ within the projected CMB-HD $2\sigma$ sensitivity.
With scalar superradiance included, $\Delta N_{\rm eff}$ falls below this threshold across the entire $(M_{\rm ini},\,a_{*,\rm ini})$ plane explored in this work (fig.~\ref{fig:contours}).
No CMB-HD iso-$\Delta N_{\rm eff}$ contour survives once superradiance is included.

Fourth, subdominant modes can matter on their own.
By setting $a_{*,\rm ini} = a_{\ast c}^{|211\rangle}$ to isolate the $|322\rangle$ mode, we find that it alone produces a non-negligible superradiant GW contribution (fig.~\ref{fig:scan}, right panel, orange dashed).

These findings have direct implications for the interpretation of $\Delta N_{\rm eff}$ measurements.
Existing constraints on PBH parameters derived under the Hawking-only assumption must be revisited if light bosons are present in the particle spectrum: superradiance reverses the monotonic relationship between initial spin and $\Delta N_{\rm eff}$, turning the highest-spin configurations from the most detectable into the most suppressed.
More broadly, the strength of the superradiant effect depends on $\alpha_{\rm ini}=M_{\rm ini}\mu$, coupling the PBH mass range to the BSM boson mass.
A measured value of $\Delta N_{\rm eff}$ can therefore constrain the combination of PBH and BSM parameters jointly: since the superradiant suppression depends on $\alpha_{\rm ini}$, a single measurement constrains a curve in the $(M_{\rm ini},\mu)$ plane rather than either parameter individually.
Conversely, a BSM model that predicts a specific boson mass $\mu$ directly determines the expected $\Delta N_{\rm eff}$ as a function of $M_{\rm ini}$, setting the target sensitivity required to probe that scenario with future CMB observations.

It is worth noting that GWs from Hawking radiation and superradiance also contribute to the stochastic gravitational wave background (SGWB).
However, the characteristic frequencies of these GWs---obtained from the PBH Hawking temperature and the BSM boson mass, respectively---are extremely high, far above the sensitivity bands of current and planned GW detectors~\cite{LIGOScientific:2014pky,LISA:2017pwj,Luo:2021qji,TianQin:2015yph,Hobbs:2009yy}. 
Consequently, $\Delta N_{\rm eff}$ provides the most stringent observational constraint on this gravitational radiation.
The precision calculation performed in this work is therefore necessary for meaningful comparison with future CMB measurements.

Several assumptions in this work point to natural extensions.
We have adopted a monochromatic PBH mass function; an extended distribution (e.g., log-normal) would assign different $\alpha$ values to different mass bins, and the integrated $\Delta N_{\rm eff}$ could develop a richer structure.
We have considered a single BSM boson species at a time; multiple species with different masses would introduce competing superradiant modes that further redistribute the angular momentum budget.
We have assumed $\beta \gg \beta_c$, which enables a temporary PBH-dominated epoch (PBH reheating scenario); for $\beta < \beta_c$, the absence of an early matter-dominated era could weaken the redshift dilution that suppresses $\Delta N_{\rm eff}$, but whether this leads to a larger $\Delta N_{\rm eff}$ requires further study.
We have also restricted the pre-BBN expansion history to standard radiation domination; non-standard cosmologies can reshape the redshifted graviton spectrum and the resulting $\Delta N_{\rm eff}$~\cite{Ireland:2023avg}.
Our analytic expressions for the superradiant growth rates are valid for $\alpha\lesssim\mathcal{O}(0.1)$; extending the analysis to larger couplings requires fully numerical treatments of the Teukolsky equation~\cite{Dolan:2007mj,East:2017ovw}.
Finally, we have not included non-linear boson self-interactions, which can trigger bosenova collapses~\cite{Arvanitaki:2010sy,Yoshino:2015nsa} at large cloud occupation numbers and alter the GW emission profile.
Each of these directions would refine the quantitative predictions presented here, but the qualitative conclusion---that superradiance is a significant and often dominant correction to the DR yield from Kerr PBHs---is expected to persist.

\acknowledgments
The authors would like to thank Prof.\ Shou-Shan Bao and Prof.\ Hong Zhang for helpful discussions.
This work is supported by the National Natural Science Foundation of China (Grants Nos.\ 12533001, 12575049, 12473001, and 12447105), the National SKA Program of China (Grants Nos.\ 2022SKA0110200 and 2022SKA0110203), the China Manned Space Program (Grant No.~CMS-CSST-2025-A02), and the 111 Project (Grant No.\ B16009). 
C.~Zhang is supported by the Joint Fund of Natural Science Foundation of Liaoning Province (Grant No.\ 2023-MSBA-067) and the Fundamental Research Funds for the Central Universities (Grant No.\ N2405011).

\bibliographystyle{JHEP}
\bibliography{main.bib}

\end{document}